\documentclass[pre,reprint,showpacs,floatfix,aps
twocolumn,
]{revtex4-1}
\usepackage{amsfonts}
\usepackage{amsmath,amssymb}
\usepackage{cancel,verbatim}
\usepackage{graphicx,float}
\usepackage{color}
\usepackage{bm}

\usepackage{epsfig}

\newcommand{\beq}{\begin{equation}}  
\newcommand{\eeq}{\end{equation}}  
\newcommand{\bea}{\begin{eqnarray}}  
\newcommand{\eea}{\end{eqnarray}}  
\newcommand{\nueff}{\hat{\nu}}
\newcommand{\eps}{\varepsilon}
\newcommand{\epsin}{\varepsilon_{\rm IN}}
\renewcommand{\vec}[1]{\bm{#1}}
\newcommand{\fvec}[1]{\hat{\bm{#1}}}
\newcommand{\fomega}{\hat{\omega}}
\newcommand{\Els}{E_{\rm LS}}
\newcommand{\Ess}{E_{\rm SS}}
\newcommand{\Ein}{E_{\rm IN}}

\renewcommand{\vec}[1]{\bm{#1}}

\newcommand{\vx}{{\vec{x}}}

\newcommand{\vu}{{\vec{u}}}

\newcommand{\vp}{{\vec{p}}}

\newcommand{\mcm}[1]{{\textcolor{black}{#1}}}
\newcommand{\blue}[1]{{\textcolor{black}{#1}}}
\newcommand{\gb}[1]{{\textcolor{black}{#1}}}

\newcommand{\Rey}{\mbox{\textit{Re}}}

\begin{document}

\title{Condensate formation and multiscale dynamics in two-dimensional active suspensions}
\author{Moritz Linkmann$^1$, M. Cristina Marchetti$^2$, Guido Boffetta$^3$, and Bruno Eckhardt$^1$}
\affiliation{
$^1$Fachbereich Physik, Philipps-Universit\"at Marburg,  D-35032 Marburg, Germany \\
$^2$Department of Physics, University of California, Santa Barbara, CA, 93106, USA \\
$^3$Dipartimento di Fisica and INFN, Universit{\`a} di Torino, via P. Giuria 1, 10125 Torino, Italy
}
\date{\today}

\begin{abstract}

The collective effects of microswimmers in active suspensions result in {\em
	active turbulence}, a spatiotemporally chaotic dynamics at mesoscale,
	which is characterized by the presence of vortices and jets at scales
	much larger than the characteristic size of the individual active
	constituents.  To describe \mcm{this} 
	dynamics, Navier-Stokes-based
	one-fluid models driven by small-scale forces have been proposed.
	Here, we provide a justification of such models for the case of dense
	suspensions in two dimensions (2d). We subsequently carry out an
	in-depth numerical study of the properties of one-fluid models as a
	function of the active driving in view of possible transition scenarios
	from active turbulence to large-scale pattern, \mcm{referred to as} 
	condensate,
	formation induced by the classical inverse energy cascade in Newtonian
	2d turbulence.  Using a one-fluid model it was recently shown (Linkmann
	{\em et al.}, Phys. Rev. Lett. (in press)) that two-dimensional active
	suspensions support two non-equilibrium steady states, one with a
	condensate and one without, which are separated by a subcritical
	transition.  Here, we report further details on this transition such as
	hysteresis and discuss a low-dimensional model that describes the main
	features of the transition through nonlocal-in-scale coupling between
	the small-scale driving and the condensate. 

\end{abstract}

\pacs{47.52.+j; 05.40.Jc}
\maketitle

\section{Introduction}
\label{sec:introduction}
Active suspensions consist of self-propelled constituents, e.g. bacteria
such as {\em Bacillus subtilis} and {\em Escherichia coli} \cite{Wu2000,Dombrowski04},
chemically driven colloids \cite{Bricard13} 
or active nematics \cite{Sanchez12,Zhou14,Giomi15}
that move  in a solvent liquid, most often water. Their collective motion 
results in complex patterns on many scales, 
and shows different phases of coherence and self-organization 
such as swarming, cluster formation, jets and vortices
\cite{Dombrowski04,Sokolov07,Cisneros07,Wolgemuth08,Wensink12,Dunkel13PRL,Gachelin14}, 
and, eventually, {\em active} or {\em bacterial turbulence} \cite{Dombrowski04}.  
The latter is a state characterized by 
spatio-temporal chaotic dynamics reminscent of vortex patterns in turbulent flows.
The analogy is not complete, though, since 
Newtonian turbulence is a multiscale phenomenon associated with and
dominated by 
dynamics in an inertial range of scales. 
Since dissipative effects are negligible in the inertial range, the rate of energy 
transfer across inertial ranges is constant, and is one of the determining 
features of 
the well-known energy cascade \cite{Frisch95}. Thus far, the
states that have been described as bacterial turbulence do not have 
an inertial range.

Active and Newtonian turbulence usually occur in different regions of parameter space.  
With Reynolds numbers $\Rey= U L / \nu$ based on typical velocities $U$, lengths $L$
and the viscosity $\nu$ of the liquid, one finds 
turbulence occurs in pipes and other flows for Reynolds numbers around 2000 
\cite{Landau59,Lautrup2005,Avila11}, 
while the mesoscale vortices observed in bacterial suspensions 
\cite{Dombrowski04} are associated with a Reynolds number of $O(10^{-3}-10^{-2})$, 
far from the inertial dynamics of Newtonian turbulence.  However, 
rheological measurements of the effective viscosity have shown 
that the active motion of the constituents can reduce the effective viscosity
by about an order of magnitude compared to the solvent viscosity
\cite{Hatwalne04,Liverpool06,Sokolov09,Gachelin13,Lopez15,Marchetti15}. 
Multiscale states at Reynolds number around 30 have been reported for larger 
microswimmers such as magnetic rotors \cite{Kokot2017}. That is, under 
favorable conditions active suspensions can reach parameter ranges
where inertial effects will influence the dynamics, and where a 
a transition from active to inertial turbulence could be achieved.

The effects of inertia are particularly intriguing in two-dimensional and quasi-two-dimensional
suspensions, as 
kinetic energy is transferred from small to large scales in 2d turbulence, 
eventually resulting in the accumulation of energy at the largest length scales 
\cite{Kraichnan67,Hossain83,Smith93,Alexakis18}. This phenomenon can be viewed in analogy
to Bose-Einstein condensation, which is why the concentration of energy on the largest scales
is called the formation of a condensate.

Full models for the dynamics of active suspensions require equations for the velocity field
and the swimmers, with suitable couplings between them \cite{Liverpool08}. Since our focus is on
the inertial effects in the flow fields, it is advantageous to eliminate the bacteria and to use
equations for the flow fields. Such one-fluid 
models of active suspensions have recently been proposed,
\cite{Wensink12,Slomka15EPJST} and have already led to a number of numerical investigations into the 
nonlinear dynamics of active suspensions that have revealed new phenomena, such as nonuniversality 
of spectral exponents \cite{Bratanov15}, mirror-symmetry breaking \cite{Slomka17PNAS}, 
or the formation of vortex lattices \cite{James18b}. 
Hints of condensation and multiscaling have been also been observed 
\cite{Oza2016,Mickelin2018}, 
but the actual formation of sizeable condensates and the connection between 2d active and Newtonian 
turbulence have not been explored systematically. 
Using a variant of these one-fluid models we have recently shown that strong condensates can form in active
suspensions, and they do so through a subcritical transition \cite{Linkmann19a}. We here provide 
further results on this transition and on the multiscale dynamics of dense active suspensions.   

This paper is organised as follows. We begin with a general discussion of continuum models for active 
suspensions in Sec.~\ref{sec:models}, including a justification of Navier-Stokes-based 
one-fluid models for dense suspensions in 2d. Section \ref{sec:numerics} contains a description 
of the datasets collected in direct numerical simulations (DNS),
followed by a discussion of the general features 
of multiscale dynamics and large-scale pattern 
formation in one-fluid models of active suspensions in Sec.~\ref{sec:model-dynamics}.  
The subcritical transition to condensate formation is described in detail in Sec.~\ref{sec:transition}
and Sec.~\ref{sec:four-scale-model} introduces a low-dimensional model that captures the 
qualitative features of the transition through a nonlocal-in-scale coupling between the condensate 
and the driven scales. 
We summarize our results in Sec.~\ref{sec:conclusions}.

\section{Models describing bacterial suspensions}
\label{sec:models}
Active suspensions consist of swimmers immersed in a fluid.
Models for such suspension have to capture the fluid and the motion of
the constituents, which, in a continuum description, leads to 
a two-fluid approach, where the solvent flow and the coarse-grained 
motion of the microswimmers are treated as separate but interacting 
quantities. 
In order to simplify the models, 
one-fluid descriptions leading to Navier-Stokes-like equations have been proposed 
\cite{Wensink12,Dunkel13PRL,Dunkel13NJP,Slomka15EPJST,Slomka17PNAS}. 
These models usually belong to one of two categories: 
(i) Bacterial flow models, where the motion of the solvent is 
eliminated in favor of the bacterial motion 
\cite{Wensink12,Dunkel13PRL,Dunkel13NJP}, or,  
(ii) solvent flow models, where a linear relation between 
a force on the solvent and 
the coarse grained 
motion of the bacteria is postulated \cite{Slomka15EPJST,Slomka17PNAS}. 
In both cases the resulting velocity field is assumed to be divergence-free, which limits
the applicability of these models to very dense suspensions where fluctuations
in the bacterial density can be neglected \cite{Wensink12}. 
In the next subsections, we motivate
a single-equation solvent flow model in two dimensions 
from the general two-fluid approach.  

\subsection{Justification of effective models in 2d}
\label{sec:justification}
At the continuum level, an active bacterial suspension can described by 
equations for the total density
$\rho$ and flow velocity $\vu$ of the suspension, the concentration $c$ of
bacteria and the coarse-grained polarization field $\vp$, which plays the dual
role of the bacterial velocity and an order parameter for collective phenomena 
in the bacteria. 
We assume that the suspension is
incompressible, i.e., $\dot\rho=0$, implying $\nabla\cdot\vu=0$, and that the
bacterial concentration is constant, resulting in 
$\nabla\cdot\vp=0$.
This then leaves two coupled equations \cite{Liverpool08}, 
given by
\begin{align}
        \label{eq:evol-u}
        \partial_t \vu + \vu \cdot \nabla \vu & = - \nabla \Pi + \nabla \cdot \sigma^a + \nu \Delta \vu \ ,  \\
        \label{eq:evol-n}
	\partial_t \vp + \vu \cdot \nabla \vp & = - \nabla \Pi'(|\vp|,c) -\lambda_1 \vp \cdot \nabla \vp+\frac{1}{2}\bm\omega\times\vp 
	\nonumber \\
	& \quad + \lambda\bm{D}\cdot\vp + \frac{1}{\gamma_F}\bm{h} \ ,
\end{align}
where 
$\Pi$ is the
pressure (divided by the total density) which ensures incompressibility of the velocity field, 
$\sigma^a$ the active stress that couples bacteria and flow (and that will be discussed further below),
$\Pi'$ is an effective pressure term that depends on the bacterial concentration and 
the polarization, 
$\bm\omega=\nabla\times\vu$ is the vorticity,
$\bm{D}=\frac12\left[\nabla\vu+(\nabla\vu)^T\right]$ the \mcm{rate of strain} 
tensor, and 
$\nu$ the kinematic viscosity of the solvent. 
The parameters $\lambda_1$ and $\lambda$ capture advective
and flow alignment, and $\mcm{\gamma_F}$ is a rotational viscosity. 

The molecular field $\bm{h}$ can be
obtained from a free energy $F$ for a polar fluid, modelled similar to a liquid crystal, 
as the derivative, $\bm{h}=-\delta F / \delta\vp$, where
\beq
\label{eq:free-energy}
F=\int \left[\frac{\alpha_F}{2}\vp^2+\frac{\beta_F}{4}\vp^4+\frac{K}{2}(\nabla\vp)^2\right] d \vx \ ,
\eeq
with $K$ the liquid crystalline stiffness in a one-elastic constant
approximation and $\alpha_F$ and $\beta_F$ the parameters which determine the onset of
a polarized state for $\alpha_F<0$. 
Note that we have neglected in both equations passive liquid-crystalline stresses of
higher order in gradients of the polarization. A derivation of Eq.~\eqref{eq:evol-n}
can be found, \mcm{for instance}, in Ref.~\cite{Liverpool08}.

The feedback of the active swimmers on the flow is contained in the 
stress tensors $\sigma^a$, which results from the active dipolar forces 
exerted on the solvent 
by the microswimmers \cite{Simha2002}. 
On length scales large compared to the size of swimmers, it can be expressed through a gradient expansion, 
with leading order term
\beq
\sigma^{a(0)}_{ij} = \alpha \left(p_i p_j-\frac{1}{3}\delta_{ij} \vp^2\right) + O(\nabla) \ ,
\label{eq:sigmaa}
\eeq
where $\alpha$ is a parameter known as activity, \mcm{that} 
depends on the concentration of microswimmers, 
their typical swimming speed and the type of swimmer. The symbol  $O(\nabla)$ indicates higher-order 
terms that contain gradients of the polarization field.
The contribution of the diagonal term $-\frac{1}{3}\delta_{ij} \vp^2$ 
can be absorbed in the pressure gradient in eq.~\eqref{eq:evol-u}.

Note that the leading-order contribution to the active stress given in eq.~\eqref{eq:sigmaa} has nematic
rather than polar symmetry, as it is parity-invariant. Indeed, 
an active stress with purely polar
symmetry arises first in terms containing 
gradients~\cite{Giomi2008}, and is given by
\beq
\sigma^{a(1)}_{ij} = \beta \left(\partial_i p_j + \partial_j p_i\right) \ ,
\label{eq:sigmaa1}
\eeq
where $\beta$ is another activity parameter \blue{that depends, amongst other quantities, on the direction of the 
polarization field with respect to the swimming direction \cite{Liverpool08,Baskaran09}}. 

In most studies of active suspensions, the fluid flow $\vu$ is slaved to the
polarization field $\vp$, resulting in the Toner-Tu model for the dynamics of
the polarization/bacterial velocity \cite{Wensink12}.  
In contrast, we here 
wish to eliminate $\vp$ in favor of $\vu$ in order to obtain an equation for active flows, 
as done for instance in Ref.~\cite{Slomka17PNAS}.

In order to derive such a single-equation model, one has to
solve the equation for $\vp$ and substitute the solution into the equation for $\vu$.
Even though the
nonlinearities in 
\eqref{eq:evol-n} 
make it difficult to obtain an analytical
solution, such an approach will generally give a functional relation between $\vp$ and $\vu$.
In what follows we show how the 2d solvent model can be obtained as the leading-order
contribution for the case of a linear,
though not necessarily local, relation between $\vp$ and $\vu$, of the form
\beq
\label{eq:p-u-general}
p_i[\vu](\vx,t) = (G_{ij}*u_j)(\vx,t) \ ,
\eeq
where $G_{ij}$ is a kernel that depends on the details of the system
and $*$ denotes a convolution. The derivation follows similar steps as
in active scalar advection in geophysical flows \cite{Constantin1998,Celani2004}.

In 2d, incompressibility of the fields reduces the number of degrees of
freedom of each vector field from two to one, usually given by the out-of-plane vorticities
$\omega(x,y)=\bm{\hat{z}}\cdot(\nabla \times \vu(x,y))$ and $m(x,y)=\bm{\hat{z}}\cdot(\nabla \times \vp(x,y))$ 
of the respective fields, \blue{where $\fvec{z}$ 
is a unit vector in the $z$-direction.} 
Equation (\ref{eq:p-u-general}) then 
becomes a scalar relation,
\beq
\label{eq:m-omega}
m[\omega](\vx,t) = (G*\omega)(\vx,t) \ .
\eeq
In a dense bacterial suspension, hydrodynamic interactions are screened and the relation between $m$ and $\omega$ is expected to be
local. We can then assume $G$ to be a sharply peaked
function, for instance proportional to a narrow spherically symmetric 2d Gaussian
\beq
G(\vx) = \frac{A}{\pi a} e^{-|\vx|^2/a^2} \ ,
\eeq
with shape parameter $a>0$ and constant amplitude $A$. 
Expanding the Fourier transform of the Gaussian in terms of its shape parameter around zero
leads to an expansion of eq.~\eqref{eq:m-omega} of the form
\beq
\label{eq:m}
m[\omega](\vx,t) =  A\omega(\vx,t) + A \frac{a^2}{4} \Delta \omega(\vx,t) + O((a^2\Delta)^2) \ ,
\eeq
where $\Delta$ the Laplace operator.
Since $\nabla \times \omega \bm{\hat{z}} = - \Delta \vu$ and similarly for $\vp$ and $m$, we obtain
\beq
\label{eq:p-u}
\vp = A\vu(\vx,t) + A \frac{a^2}{4} \Delta \vu(\vx,t) + O((a^2\Delta)^2) \ .
\eeq
Inserting Eq.~\eqref{eq:p-u} into Eq.~\eqref{eq:sigmaa} for the zeroth-order term yields
\begin{align}
	\sigma^{a(0)}_{ij} & = \alpha A^2 \left( u_i u_j-\frac{1}{3}\delta_{ij} \vu^2  \right) 
		   \gb{+ \alpha A^2 {a^2 \over 4} \left( u_i \Delta u_j + u_j \Delta u_i \right)} \nonumber \\
		   & + \alpha A^2 \frac{a^4}{16} \left(\Delta u_i \Delta u_j-\frac{1}{3}\delta_{ij} (\Delta \vu)^2 \right) 
		    + O((a^2\Delta)^4) \ , 
\label{eq:sigmaa-u}		    
\end{align}
resulting in additional quadratic nonlinearities in Eq.~\eqref{eq:evol-u}, 
\mcm{some of which break Galilean invariance}. 
The first term in eq.~\eqref{eq:sigmaa-u} leads to a renormalisation of 
the Navier-Stokes nonlinearity and hence to a different Reynolds number. Since the 
sign of $\alpha$ depends on the type of swimmer with $\alpha < 0$ for pullers and $\alpha > 0$ for 
pushers, the renormalisation of the Reynolds number depends on the type of microswimmers. 
The second term can be subsumed into the pressure gradient, while remaining terms which are of 
higher order in the gradients contribute to a redistribution of kinetic energy mostly at small scales.   
Since all terms conserve the mean kinetic energy and hence do not result in a net energy input, we neglect the 
additional small-scale nonlinearities, \mcm{thereby ensuring Galilean invariance.}
The energy input from the microswimmers hence has to originate from \gb{the first-order term in the gradient expansion of 
the active stresses given in Eq.~\eqref{eq:sigmaa1}}. 
Substituting Eq.~\eqref{eq:p-u} in Eq.~\eqref{eq:sigmaa1} results in 
\begin{align}
	\sigma^{a(1)}_{ij} & = \beta A\left(1+ \frac{a^2}{4} \Delta + \frac{a^4}{32} \Delta^2 \right) 
	 \left(\partial_i u_j + \partial_j u_i\right) \nonumber \\
	                    & + O((a^2\Delta)^3) \ ,
\label{eq:sigmaa1-u}		    
\end{align}
\gb{where terms up to order $\Delta^2$ from Eq.~\eqref{eq:p-u} have been included based on stability considerations, }
and the structure of $\nabla \cdot \sigma^{a(1)}$ takes on the form of the effective viscosity 
previously proposed 
by  S{\l}omka and Dunkel \cite{Slomka15EPJST}, provided $\beta > 0$. The latter is the case if 
$\vp$ is chosen to point along the swimming direction and does not depend on the type of microswimmer 
\cite{Baskaran09}. If we choose $\vp$ to point against the swimming direction, 
then $A$ should be negative. That is, the product $\beta A$ is always positive. 
In what follows we choose $A>0$ such that $\vp$ points into the same direction as the solvent flow.
After the 
rescaling
\beq
t \to t\sqrt{1-\alpha A}, \quad \vu \to \vu \sqrt{1-\alpha A}, 
\eeq
the resulting two-dimensional one-fluid model reads
\begin{align}
\label{eq:sf-model}
	\partial_t \vu + \vu \cdot \nabla \vu & = - \nabla \Pi + \Gamma \left( \Gamma_0 + \Gamma_2 \Delta + \frac{\Gamma_2^2}{2} \Delta^2\right) \Delta \vu \ ,  
	\nonumber \\
	\nabla \cdot \vu & = 0 \ ,
\end{align}
where 
\beq
\Gamma  = \frac{\beta A}{\sqrt{1-\alpha A}}, \quad
\Gamma_0  = 1+\frac{\nu}{\beta A}, \quad
\Gamma_2  = \frac{a^2}{4} \ . 
\eeq

Equation \eqref{eq:sf-model} relies upon two main assumptions: (i) $\vu$ and
$\vp$ are divergence-free, i.e. the bacterial concentration must be constant
and density fluctuations negligible; (ii) the system must be two-dimensional, as
the reduction to a one-dimensional problem resulting in
Eq.~\eqref{eq:m-omega} is not justified otherwise. 
Specifically, in three dimensions there is no {\em
a-priori} reason to set $G_{ij} = G \delta_{ij}$ in Eq.~\eqref{eq:p-u-general}.
In summary, Eq.~\eqref{eq:sf-model} is applicable to dense suspensions of
microswimmers in very thin layers, where a 2d-approximation is justified. 
\mcm{We note that friction with a substrate has been neglected, however, 
the corresponding term can easily be added}.

In this context, the original introduction of the solvent model by S{\l}omka
and Dunkel corresponds to setting $G(\vx) \sim \delta(\vx)$ and using certain
higher-order terms in the gradient expansion of the active stresses.  The
former amounts to assuming that
the polarization and solvent velocity fields are related only locally and the
latter introduces additional parameters.
Physically, {\em locally} means on scales smaller than that of the mesoscale
vortices.  Here, we obtain a very similar model from a long-range relation
between the fields, which is more appropriate in a hydrodynamic context. 

\blue{
Similarly, the bacterial flow model introduced in Ref.~\cite{Wensink12} can be obtained by 
formally solving Eq.~\eqref{eq:evol-u} to obtain $\vu[\vp]$, or by neglecting $\vu$ altogether. 
}
Both solvent and bacterial flow models reproduce the experimentally observed
spatiotemporally chaotic dynamics characteristic of active matter turbulence 
\cite{Wensink12,Dunkel13PRL,Slomka17PNAS} and 
have been used extensively in  investigations thereof 
\cite{Dunkel13NJP,Bratanov15,Oza2016,Slomka15EPJST,Srivastava16,Putzig16,Slomka17PNAS,Slomka17PRF,
Doostmohammadi17,James18a,James18b,Mickelin2018}.
They differ in the choice of fields in which the model is expressed, with the 
consequence that terms originating from the free energy are not explicitly present in 
the solvent model. 

\blue{
The solvent models resemble the Navier-Stokes equations in their structure
and have the key ingredient for an inertial range that is typical of normal turbulence:
in the absence of forcing and dissipation, the nonlinear terms in the equation preserve the mean kinetic 
energy $\langle |\vu|^2\rangle$. 
\mcm{We note
that in general Galilean invariance is broken by active corrections coming from
Eq.~\eqref{eq:sigmaa-u} to the advective nonlinearity in Eq.~\eqref{eq:sf-model}
that have been neglected here.}
The effects of the active particles are thus
concentrated in the effective viscosity in Eq.~\eqref{eq:sf-model}, and we will focus
on two variants of the models and discuss
similarities and differences to results in the literature that were obtained
with the bacterial flow model.}

\subsection{Polynomial effective viscosity}
\label{sec:poly-model}
The solvent model introduced by S{\l}omka and Dunkel \cite{Slomka15EPJST}
has a stress tensor in Eq.~\eqref{eq:evol-u} given by a polynomial gradient
expansion 
\beq
\label{eq:stress-poly}
\sigma_{ij} = \left(\Gamma_0 + \Gamma_2 \Delta + \Gamma_4 \Delta^2 \right)
 \left(\partial_i u_j + \partial_j u_i\right) \, ,
\eeq
and results in a continuous effective viscosity 
\beq
\label{eq:nuk-poly}
\nueff(k) = \Gamma_0 - \Gamma_2 k^2 + \Gamma_4 k^4 \, .
\eeq
where $\hat{\cdot}$ denotes the Fourier transform. 
In what follows, we will therefore refer to the combination of Eqs.~\eqref{eq:evol-u} 
and \eqref{eq:stress-poly} as the polynomial effective viscosity (PEV) model.

If $\Gamma_2<0$, then Eq.~\eqref{eq:nuk-poly} is a combination of normal and hyperviscosity
and all terms dissipate energy. If, in contrast, $\Gamma_2>0$,  
\blue{there is a wave number interval where $\hat \nu(k) < 0$, resulting in a linear amplification
of the Fourier modes in that wave number interval.}
The interplay between this instability and the Navier-Stokes nonlinearity 
drives spatiotemporal dynamics that for certain values of $\Gamma_i$
resemble the experimental observations \cite{Slomka17PNAS}. 

By completing the square, the wave number form of the effective viscosity
can be written as
\beq
\label{eq:nu_k-poly}
\nueff(k) = \Gamma_0 + \Gamma_4 \left( \left(k^2-k_{\rm f}^2 \right)^2 - k_{\rm f}^4 \right)
\eeq
with
\beq
k_{\rm f}^2 = \Gamma_2/(2\Gamma_4)
\eeq
 the wave number of the
minimum in the viscosity (which is real only for $\Gamma_2>0$).
With the normalization of wave numbers to $k_{\rm f}$, i.e. $\tilde k=k/k_{\rm f}$,
the effective viscosity can be written
\beq
\nueff(k) = \Gamma_0 \left(1 + \gamma \left( \left(\tilde k^2-1 \right)^2 - 1\right) \right)
\label{scaled_nueff}
\eeq
where
\beq
\gamma = \Gamma_4 k_{\rm f}^4 /\Gamma_0 = \frac{\Gamma_2^2}{4\Gamma_0 \Gamma_4}
= \frac{(\Gamma_2/\Gamma_0)^2}{4(\Gamma_4/\Gamma_0)}
\eeq
\blue{is the one remaining parameter that controls the forcing}.

\begin{figure}
	\includegraphics[width=0.45\textwidth]{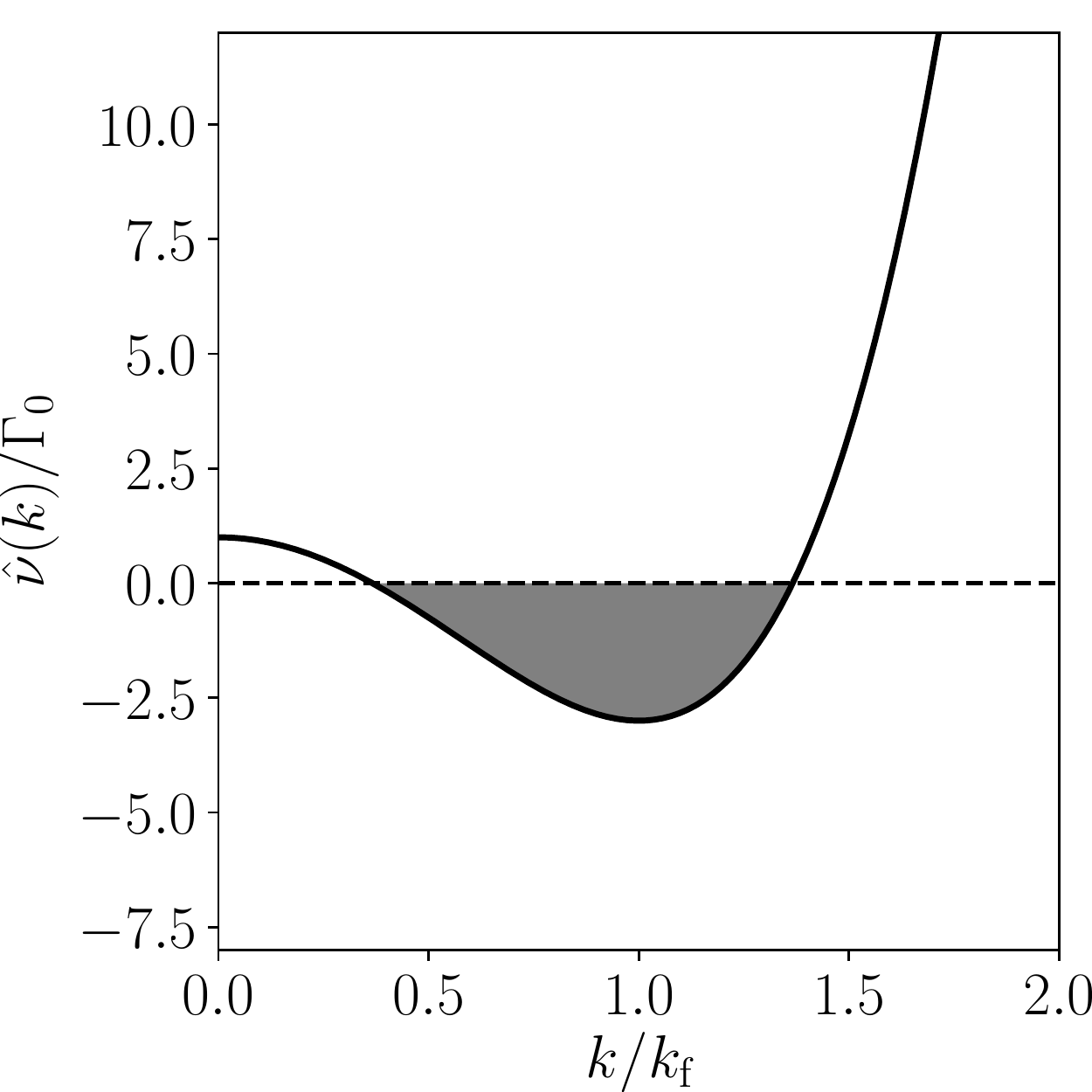} 
	\includegraphics[width=0.45\textwidth]{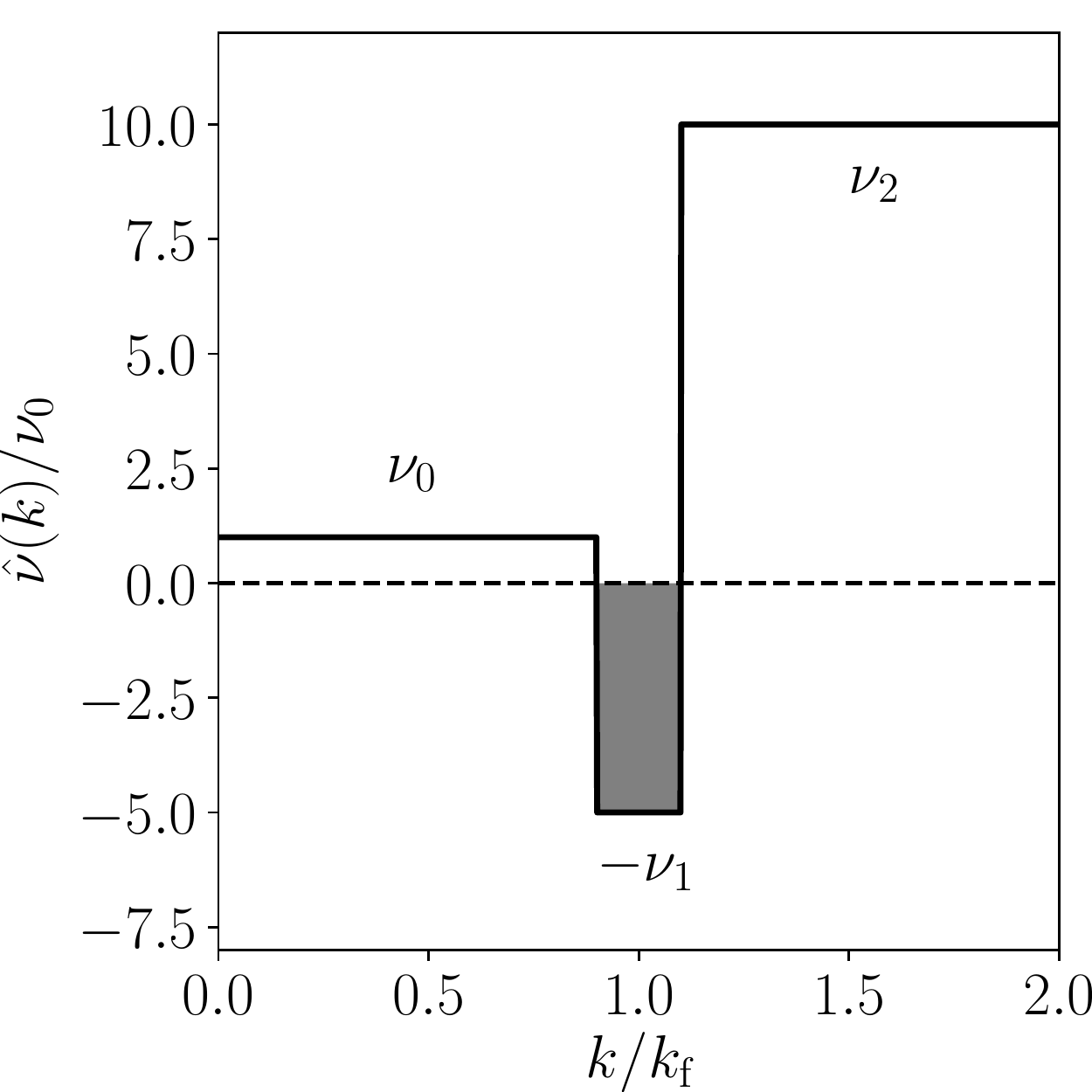} 
	\caption{
      Effective viscosity for PEV and PCV models. Top: PEV $\nueff/\Gamma_0$ vs $k/k_{\rm f}$,  
      bottom: PCV $\nueff(k)/\nu_0$ vs $k/k_{\rm f}$ . 
      The gray-shaded area corresponds to the interval $[k_{\rm min}, k_{\rm max}]$, 
      where the amplification occurs.
      }
  \label{fig:nueff}
\end{figure}

The scaled effective viscosity $\nueff$ has two parameters, $\Gamma_0$, which
sets the scale for the viscosity, and $\gamma$, which is a measure for both
the amplification and the range of wave numbers that are forced, as we will
now discuss. The effective viscosity attains its minimum at $\tilde k^2=1$,
where
\beq
\nueff(\tilde k=1) =\Gamma_0 (1-\gamma)
\eeq
Clearly, $\nueff$ can become negative, and hence forcing rather than dissipating,
for $\gamma_c>1$ only. The range of wavenumbers over which it is forcing
is given by
\beq
\label{eq:interval_k-poly}
\tilde k_{\rm min}^2 = 1-\sqrt{1-\frac{1}{\gamma}} < \tilde k^2 < 1+ \sqrt{1-\frac{1}{\gamma}} 
= \tilde k_{\rm max}^2 \ ,
\eeq
and varies with $\gamma$.
A sketch of $\nueff(k)$ for the PEV model is provided in the top panel of Fig.~\ref{fig:nueff},
the gray-shaded area indicating the wavenumber interval where amplification occurs, $\nueff(k) < 0$.
The upper end of the interval approaches $2$ for $\gamma\rightarrow\infty$,
showing that there will be no forcing on smaller wavelengths, whereas
the lower end of the interval approaches $0$, indicating that the driving
band extends to ever lower wave numbers and thus larger scales in this limit.

Since the effective viscosity is measured in units of $\Gamma_0$
and since the length scale has been fixed as
$L_{\rm f} = \pi / k_{\rm f}$
all scales in the momentum equation are set: specifically, time is
measured in units of $L_{\rm f}^2/\Gamma_0$ and velocity in units
of $\Gamma_0/L_{\rm f}$. Introducing that scale, Eq.~\eqref{eq:evol-u} contains
a single parameter $\gamma$, 
with the stress tensor is given by
\beq
\label{eq:pev-single-param}
\sigma_{ij} = \left(1 +\gamma  \left( (1+\Delta)^2-1\right)\right)
 \left(\partial_i u_j + \partial_j u_i\right) \ .
\eeq
Variations in $\gamma$ should therefore
give rise to different dynamics. S{\l}omka and Dunkel \cite{Slomka15EPJST} discuss
statistically steady states for several values of their control parameters $\Gamma_0$, 
$\Gamma_2$, and $\Gamma_4$,  that is, in our notation for different values of $\gamma$ 
and corresponding driving scales and amplitudes. Some of these states were multiscale
with energy spectra reminiscent of fully developed 2d turbulence \cite{Slomka15EPJST,Mickelin2018}, 
and small condensates were observed for certain parameter values \cite{Mickelin2018}. 
\blue{One can then expect }
that stronger large-scale structures may form for more intense driving, but since
$\gamma$ controls not only the strength of the forcing but also the width 
and the location of the driving band, it is %
is difficult to see which of the effects dominate. 
This was remedied in the model used in \cite{Linkmann19a} and described next, where amplification and
driving scale can be set independently from each other.  

\subsection{Piecewise constant viscosity}
\label{sec:pcv-model}
The piecewise constant viscosity (PCV) model \cite{Linkmann19a} is a discontinuous 
approximation to the PEV model, with the 
Navier-Stokes stress tensor 
\mcm{written in terms of}
an effective viscosity given as a set of step functions in Fourier space
\beq
\label{eq:nu_k-pcv}
\hat\nu(k) =
\begin{cases}
\nu_0 > 0 \quad \text{for} \quad k < k_{\rm min} \ ,\\
- \nu_1 < 0 \quad \text{for} \quad k_{\rm min} \leqslant k \leqslant k_{\rm max} \ ,\\
\nu_2 > 0 \quad \text{for} \quad k > k_{\rm max} \ . 
\end{cases}
\eeq
The values of $\nu_i$ are chosen such that the resulting discrete form of $\nueff(k)$ resembles
the polynomial form of the PEV model. Specifically, $\nu_1$ controls the forcing, and $\nu_2 > \nu_0$ 
mimics the hyperviscous term in the PEV model. 
A sketch of $\nueff(k)$ for the PCV model is provided in the bottom panel of Fig.~\ref{fig:nueff}, 
the gray-shaded area indicating the wavenumber interval where amplification occurs, $\nueff(k) = -\nu_1 < 0$.   
As in the PEV model, $\nu_0$ sets the scale for the effective viscosity. The 
PCV model can thus be described by dimensionless parameters for amplification, 
$\nu_1/\nu_0$, and small-scale dissipation, $\nu_2/\nu_0$. 
An effective driving scale $L_{\rm f} = \pi/k_{\rm f}$ can be defined by the 
midpoint of the interval $[k_{\rm min}, k_{\rm max}]$, 
i.e. $k_{\rm f} = (k_{\rm min}+k_{\rm max})/2$. 

\begin{table*}[t]
\centering
\begin{tabular}{lcccccccccccc}
\hline
Run id	& $N$ & $\Gamma_2/\Gamma_0$ & $\Gamma_4/\Gamma_0$ & $k_{\rm min}$ & $k_{\rm max}$ & Re & $U$ & $L$ & Re$_f$ & $\eps_{\rm LS}$ & $\eps_{\rm IN}$ & $\eps_{\rm SS}$ \\
\hline
PEV-1    &  256 &   0.002  &  $7.7 \times 10^{-7}$  & 26  & 43  & 14  &  0.10 &  0.15 &  5 & 0.0012 &  0.002  & 0.0008 \\ 
PEV-2    &  256 &   0.0023 &  $9.3 \times 10^{-7}$  & 23  & 45  & 68  &  0.22 &  0.35 &  9 & 0.0044 & 0.0072  & 0.0028 \\ 
PEV-3    &  256 &   0.0025 &  $9.7 \times 10^{-7}$  & 22  & 45  & 668  & 0.66 &  1.56 & 10 & 0.0097 & 0.0155  & 0.0056 \\ 
\hline
Run id	& $N$ & $\nu_1/\nu_0$ & $\nu_2/\nu_0$ & $k_{\rm min}$ & $k_{\rm max}$ & Re & $U$ & $L$ & Re$_f$ & $\eps_{\rm LS}$ & $\eps_{\rm IN}$ & $\eps_{\rm SS}$ \\
\hline
PCV-A1$^*$   & 256 &  0.25   & 10.0 & 33  & 40  &    19    &  0.29 & 0.07  &  19  & 0.029    & 0.048  & 0.023 \\
PCV-A2$^*$   & 256 &  0.5    & 10.0 & 33  & 40  &    26    &  0.36 & 0.085 &  21  & 0.056    & 0.10   & 0.046 \\
PCV-A3$^*$   & 256 &  0.75   & 10.0 & 33  & 40  &    35    &  0.39 & 0.09  &  21  & 0.086    & 0.16   & 0.071 \\
PCV-A4$^*$   & 256 &  1.0    & 10.0 & 33  & 40  &    44    &  0.43 & 0.11  &  21  & 0.11     & 0.21   & 0.09 \\
PCV-A5$^*$   & 256 &  1.25   & 10.0 & 33  & 40  &    58    &  0.47 & 0.13  &  21  & 0.15     & 0.26   & 0.14 \\
PCV-A6$^*$   & 256 &  1.5    & 10.0 & 33  & 40  &    75    &  0.52 & 0.15  &  20  & 0.18     & 0.30   & 0.16 \\
PCV-A7$^*$   & 256 &  1.75   & 10.0 & 33  & 40  &   106    &  0.57 & 0.20  &  19  & 0.17     & 0.34   & 0.16 \\
PCV-A7a      & 256 &  1.75   & 10.0 & 33  & 40  &   106    &  0.57 & 0.20  &  20  & 0.18     & 0.32   & 0.15 \\
PCV-A8$^*$   & 256 &  2.0    & 10.0 & 33  & 40  &   212    &  0.66 & 0.35  &  19  & 0.18     & 0.36   & 0.18 \\
PCV-A8a      & 256 &  2.0    & 10.0 & 33  & 40  &   227    &  0.67 & 0.37  &  19  & 0.19     & 0.34   & 0.17 \\
PCV-A9$^*$   & 256 &  2.02   & 10.0 & 33  & 40  &   249    &  0.68 & 0.40  &  19  & 0.18     & 0.36   & 0.18 \\
PCV-A9a      & 256 &  2.02   & 10.0 & 33  & 40  &  2686    &  1.64 & 1.79  &  17  & 0.15     & 0.28   & 0.13 \\
PCV-A10$^*$  & 256 &  2.04   & 10.0 & 33  & 40  &   296    &  0.70 & 0.46  &  19  & 0.18     & 0.36   & 0.18 \\
PCV-A10a     & 256 &  2.04   & 10.0 & 33  & 40  &  2957    &  1.77 & 1.82  &  17  & 0.14     & 0.28   & 0.13 \\
PCV-A11$^*$  & 256 &  2.083  & 10.0 & 33  & 40  &  3347    &  1.95 & 1.87  &  17  & 0.13     & 0.28   & 0.15 \\
PCV-A11a     & 256 &  2.083  & 10.0 & 33  & 40  &  3270    &  1.92 & 1.85  &  17  & 0.14     & 0.28   & 0.13 \\
PCV-A12$^*$  & 256 &  2.167  & 10.0 & 33  & 40  &  3708    &  2.13 & 1.90  &  16  & 0.13     & 0.27   & 0.15 \\
PCV-A13$^*$  & 256 &  2.25   & 10.0 & 33  & 40  &  3927    &  2.24 & 1.91  &  16  & 0.13     & 0.28   & 0.15 \\
PCV-A14$^*$  & 256 &  2.5    & 10.0 & 33  & 40  &  4455    &  2.52 & 1.92  &  15  & 0.12     & 0.27   & 0.15 \\
PCV-A15$^*$  & 256 &  2.625  & 10.0 & 33  & 40  &  4636    &  2.63 & 1.92  &  14  & 0.11     & 0.26   & 0.15 \\
PCV-A16$^*$  & 256 &  2.75   & 10.0 & 33  & 40  &  4851    &  2.75 & 1.92  &  14  & 0.11     & 0.26   & 0.15 \\
PCV-A17$^*$  & 256 &  2.875  & 10.0 & 33  & 40  &  5088    &  2.89 & 1.92  &  14  & 0.11     & 0.27   & 0.16 \\
PCV-A18$^*$  & 256 &  3.0    & 10.0 & 33  & 40  &  5313    &  3.01 & 1.92  &  14  & 0.11     & 0.28   & 0.17 \\
PCV-A19$^*$  & 256 &  3.25   & 10.0 & 33  & 40  &  5793    &  3.28 & 1.92  &  14  & 0.12     & 0.29   & 0.18 \\
PCV-A20$^*$  & 256 &  3.5    & 10.0 & 33  & 40  &  6241    &  3.54 & 1.92  &  14  & 0.13     & 0.31   & 0.19 \\
PCV-A21$^*$  & 256 &  3.75   & 10.0 & 33  & 40  &  6708    &  3.80 & 1.93  &  14  & 0.13     & 0.33   & 0.20 \\
PCV-A22$^*$  & 256 &  4.0    & 10.0 & 33  & 40  &  7214    &  4.08 & 1.93  &  14  & 0.14     & 0.35   & 0.22 \\
PCV-A23$^*$  & 256 &  4.25   & 10.0 & 33  & 40  &  7723    &  4.27 & 1.93  &  14  & 0.16     & 0.38   & 0.24 \\
PCV-A24$^*$  & 256 &  4.5    & 10.0 & 33  & 40  &  8230    &  4.65 & 1.93  &  14  & 0.17     & 0.40   & 0.25 \\
PCV-A25$^*$  & 256 &  4.75   & 10.0 & 33  & 40  &  8751    &  4.95 & 1.93  &  14  & 0.18     & 0.43   & 0.27 \\
PCV-A26$^*$  & 256 &  5.0    & 10.0 & 33  & 40  &  9258    &  5.24 & 1.93  &  14  & 0.19     & 0.46   & 0.29 \\
PCV-A27      & 256 &  5.25   & 10.0 & 33  & 40  &  9690    &  5.51 & 1.92  &  15  & 0.21     & 0.53   & 0.32 \\
PCV-A28$^*$  & 256 &  5.5    & 10.0 & 33  & 40  & 10286    &  5.81 & 1.93  &  15  & 0.23     & 0.57   & 0.34 \\
PCV-A29$^*$  & 256 &  6.0    & 10.0 & 33  & 40  & 11416    &  6.44 & 1.93  &  15  & 0.27     & 0.65   & 0.39 \\
PCV-A30$^*$  & 256 &  6.5    & 10.0 & 33  & 40  & 12530    &  7.08 & 1.93  &  15  & 0.31     & 0.74   & 0.44 \\
PCV-A31$^*$  & 256 &  7.0    & 10.0 & 33  & 40  & 13677    &  7.77 & 1.93  &  16  & 0.36     & 0.84   & 0.49 \\
\hline
PCV-B1$^*$ & 1024 & 1.0      & 10.0 & 129 & 160 & 45       & 0.027 & 0.029 & 21   & 0.0001   & 0.00019 & 9 $\times 10^{-5}$ \\
PCV-B2$^*$ & 1024 & 2.0      & 10.0 & 129 & 160 & 226      & 0.041 & 0.094 & 20   & 0.00017  & 0.00033 & 0.00016            \\
PCV-B3$^*$ & 1024 & 5.0      & 10.0 & 129 & 160 & 132914   & 1.17  & 1.93  & 15   & 0.00018  & 0.00046 & 0.00026            \\
\hline
\end{tabular}
 \caption{
Parameters and observables for all simulations, with $N$ 
denoting the number of grid points in each coordinate \gb{of the simulation domain $[0,2\pi]^2$}, 
$\nu_0$, $\nu_1$ and $\nu_2$, are 
the parameters defining the PCV-model as in Eq.~\eqref{eq:nu_k-pcv}
with $\nu_0 = 0.0011$ for PCV-A and $\nu_0 = 1.7 \times 10^{-5}$ for PCV-B. 
For PEV, the model parameters in Eq.~\eqref{eq:nu_k-poly} are
$\Gamma_0 = 0.0011$, $\Gamma_2/\Gamma_0$ and $\Gamma_4/\Gamma_0$. 
The driven intervals are specified by
$k_{\rm min}$ and $k_{\rm max}$ as defined in Eq.~\eqref{eq:nu_k-pcv} for 
PCV and in Eq.~\eqref{eq:interval_k-poly} for PEV.
The Reynolds number ${\rm Re}$ is based on
the integral scale
$L=2/U^2 \int_0^\infty dk \ E(k)/k$ and the rms velocity $U$, and
Re$_f$ is the Reynolds number based on the effective driving scale $L_{\rm f}$
and the velocity in the driven range of scales,
$\eps_{\rm LS}$ the energy dissipation rate in the interval $[1,k_{\rm min})$,
$\eps_{\rm IN}$ the energy input rate in the interval $[k_{\rm min},k_{\rm max}]$, and
$\eps_{\rm SS}$ the energy dissipation rate in the interval $(k_{\rm max},2\pi/(N/3)]$.
All observables are ensemble-averaged during the statistically stationary state, 
with samples taken at intervals of one large-eddy turnover time $T=L/U$.
The asterisk indicates data from Ref.~\cite{Linkmann19a}.
}
 \label{tab:simulations}
\end{table*}

The PCV model approximates the functional form of the PEV model's effective
viscosity by a piecewise constant function, remaining faithful to the original
PEV model in an important point: The driving is proportional to the velocity
field and it is confined to a wavenumber band. This results in driving through
local-in-scale amplification in both cases, i.e. in essentially the same
physics.  That is, even though the small-scale properties of the velocity
fields obtained by the PCV and the original PEV model may differ in some
detail, the large-scale and mean properties should be similar, if not the same,
as they are dominated by the nonlinearity and not by details of how the driven
interval is specified.

\section{Direct numerical simulations}
\label{sec:numerics}
The PEV and PCV models are studied in two dimensions, using
data generated by numerical integration of the momentum equation 
in vorticity form  
\beq
\label{eq:vort-spectral}
\partial_t \fomega(\vec{k}) + 
\widehat{\left[\vec{u} \cdot \nabla\omega\right]}({\vec{k}})
= -\hat\nu(k)k^2 \fomega(\vec{k})\ ,
\eeq
where $\omega$ is the \mcm{only non-vanishing component of the} 
vorticity, $\nabla \times \vec{u}(x,y) = \omega(x,y) \fvec{z}$. 
Equation \eqref{eq:vort-spectral} is 
supplemented with either Eq.~\eqref{eq:nu_k-poly} for PEV or 
Eq.~\eqref{eq:nu_k-pcv} for the PCV model.
In all cases, we use the standard pseudospectral technique \cite{Orszag69}
on the domain $[0,2\pi]^2$ with periodic boundary conditions and 
full dealiasing by truncation following the 2/3rds rule \cite{Orszag71}. 
The simulations are initialised with random Gaussian-distributed data, or, 
in case of hysteresis calculations for the PCV model, 
with data obtained from another run at a different value of the control parameter. 

The PEV model is only investigated for a small number of 
test cases corresponding to the parameters
specified in table \ref{tab:simulations}.  
For PCV, two series of simulations 
were carried out. \blue{The first one,} 
PCV-A, consists of a parameter scan in
$\nu_1/\nu_0$  with all other parameters, i.e. $\nu_0$,
$\nu_2$, $k_{\rm min}$ and $k_{\rm max}$ held fixed. That is, only the
amplification is varied between the simulations in each PCV-A dataset. 
The three simulations of \blue{the second} series, PCV-B, 
\blue{were done} at higher resolution, 
with parameters chosen such that results can be compared with PCV-A using the scaling 
properties of the Navier Stokes equations, 
i.e. PCV-B corresponds to PCV-A in a larger simulation domain.
Parameters and observables of all runs are summarised in table \ref{tab:simulations}.

All simulations reach a statistically stationary state, 
\blue{where the total energy per unit volume fluctutates about a mean value}, 
and are subsequently
continued for at least 2000 large eddy turnover times. Prior to that, the
system evolves through a transient non-stationary stage.  Owing to the absence
of a large-scale dissipation mechanism, this can take a long time for certain
parameter regimes.  During the statistically stationary state, the velocity
fields were sampled in intervals of one large-eddy turnover time.  

\section{Model dynamics}
\label{sec:model-dynamics}

We begin our study of 
the properties of the models by tracking 
the time evolution of the total kinetic energy per unit volume, 
$E(t)$, given by the difference between input and dissipation,
\beq
\label{eq:balance-integrated}
\frac{dE}{dt} = \eps_{\rm IN}(t)-(\eps_{\rm LS}(t)+\eps_{\rm SS}(t)) \ ,
\eeq
where the input $\eps_{\rm IN}$, the large-scale dissipation $\eps_{\rm LS}$ and 
the small-scale dissipation $\eps_{\rm SS}$ are \blue{obtained by integrating
the effective viscosity over the piecewise constant intervals, i.e.}
calculated as
\begin{align}
\label{eq:input}
	\eps_{\rm IN}(t) & = - \nu_1 \int_{k_{\rm min}}^{k_{\rm max}} d k \int d\fvec{k} 
	|\fvec{u}(\vec{k},t)|^2 \ , \\
\label{eq:ls-dissipation}
	\eps_{\rm LS}(t) & =  \nu_0 \int_0^{k_{\rm min}} d k \int  d\fvec{k} 
	|\fvec{u}(\vec{k},t)|^2 \ , \\
\label{eq:ss-dissipation}
	\eps_{\rm SS}(t) & =  \nu_2 \int_{k_{\rm max}}^\infty d k \int  d\fvec{k} 
	|\fvec{u}(\vec{k},t)|^2 \ , 
\end{align}
with $\fvec{k}= \vec{k}/k$ a unit vector in direction of $\vec{k}$.
During statistically stationary evolution, mean energy input must equal mean energy dissipation, 
$\eps_{\rm IN} = \varepsilon = \eps_{\rm LS}+\eps_{\rm SS}$. The characteristics of the 
non-stationary evolution depends on the presence of an inverse energy transfer.
If an inverse cascade is present, as in fully developed 2d turbulence, it can be expected 
that $E(t)$ grows linearly in time as long as $\eps_{\rm LS}$ is negligible. This is a consequence
of the fact that 
the dynamics at the small scales are much faster than at large scales leading to 
$\eps_{\rm IN} \simeq const$ and $\eps_{\rm SS} \simeq const$, and one obtains
\beq
E(t) \simeq (\eps_{\rm IN} - \eps_{\rm SS})t \ ,
\eeq
until $\eps_{\rm LS}$ becomes sufficiently large.  

\begin{figure}[tbp]
\centering
	\includegraphics[width=\columnwidth]{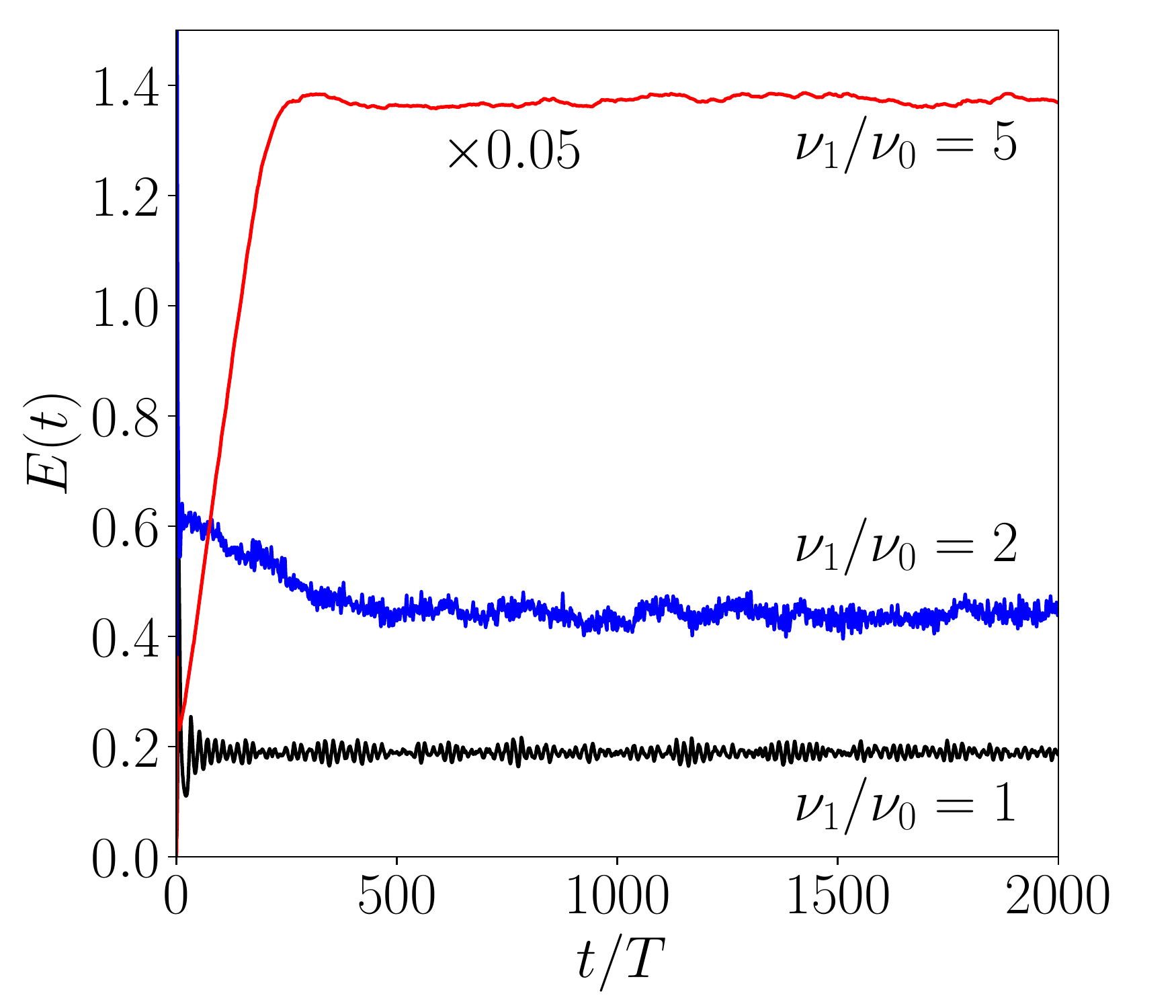} 
	\includegraphics[width=\columnwidth]{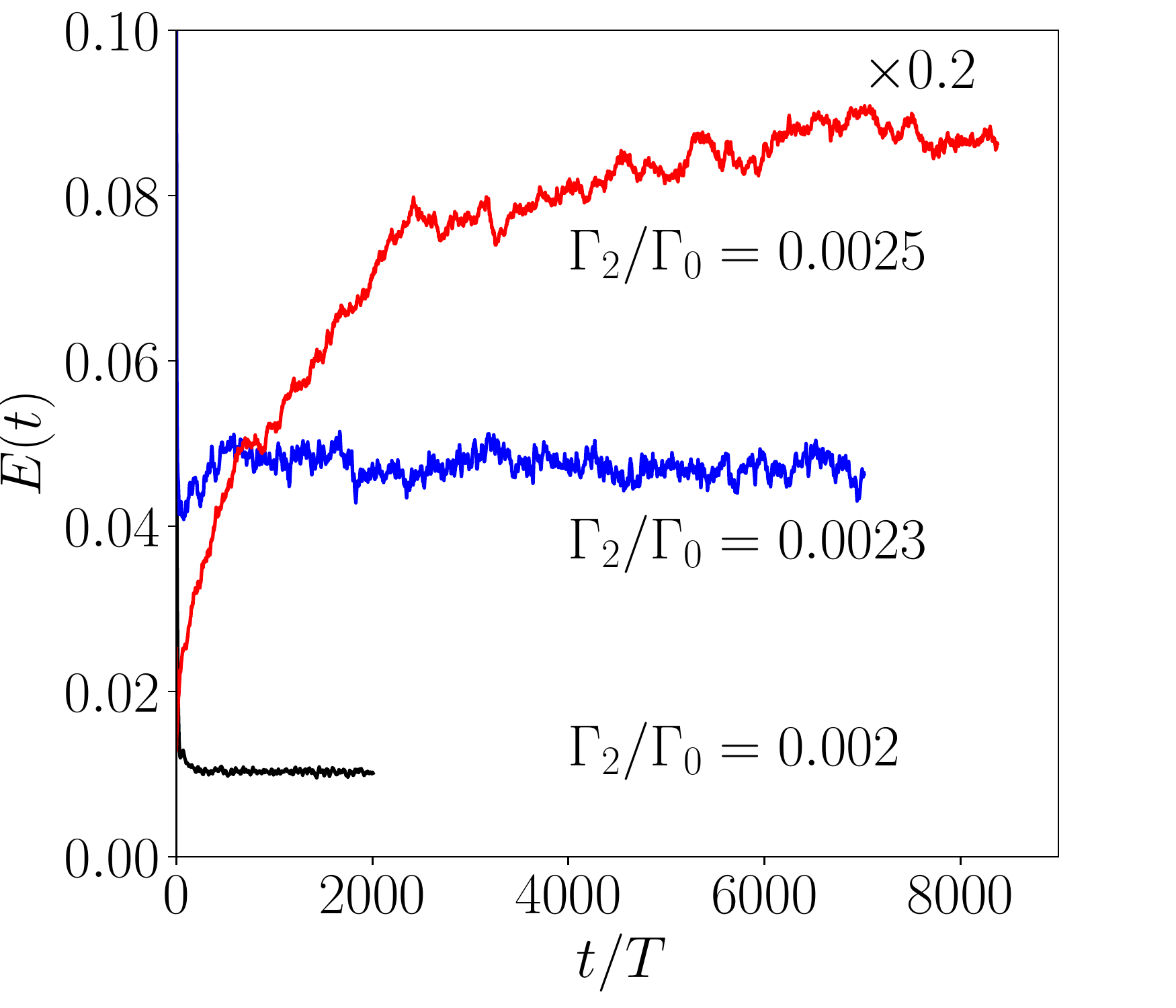} 
 \caption{
(Color online) Time evolution of the total kinetic energy per unit volume
	for three example cases for PCV (top) and PEV (bottom). 
	The energy has been divided by a factor of 20 for the PCV case $\nu_1/\nu_0 =5$
	and by a factor of 5 for the PEV case $\Gamma_2/\Gamma_0 = 0.0025$
	in order to improve the readability of the figure. 
         }
 \label{fig:time-evol}
\end{figure}

The time evolution of the total energy per unit volume, $E(t)$, is shown in 
Fig.~\ref{fig:time-evol} for representative cases.
%
In the top panel, we show the results for the runs PCV-B1, PCV-B2 and PCV-B3 with amplification factors 
$\nu_1/\nu_0 = 1$, $\nu_1/\nu_0 = 2$ and $\nu_1/\nu_0 = 5$, respectively.
The lower frame shows the corresponding results for PEV 
with parameters
$\Gamma_2/\Gamma_0 = 0.0025$,
$\Gamma_2/\Gamma_0 = 0.0023$, and
$\Gamma_2/\Gamma_0 = 0.002$, with $\Gamma_4$ chosen such that the forcing remains 
centered around $k_{\rm f}=36$.

The behavior of $E(t)$ is qualitatively similar for the two models and 
differs between the respective example cases. The two cases with low amplification, 
that is, $\nu_1/\nu_0 = 1$ and $\nu_1/\nu_0 = 2$ for PCV and 
$\Gamma_2/\Gamma_0 = 0.002$ and $\Gamma_2/\Gamma_0 = 0.0023$ for PEV
become statistically stationary and fluctuate around relatively low mean values of $E$. 
In contrast, for the cases $\nu_1/\nu_0 = 5$ for PCV and $\Gamma_2/\Gamma_0 = 0.0025$ for PEV, 
the kinetic energy grows at first linearly, which is characteristic of a
non-stationary inverse energy cascade in 2d turbulence \cite{Boffetta14ARFM}. 
This is followed by statistically stationary evolution, where $E(t)$ fluctuates about mean 
values which are an order of magnitude larger than for the aforementioned cases.
In absence of a large-scale friction term, once an inverse energy
transfer is established, statistical stationarity can only be realized
through the development of a condensate at the largest scales. 

\begin{figure*}
\centering
	\includegraphics[width=0.3\textwidth]{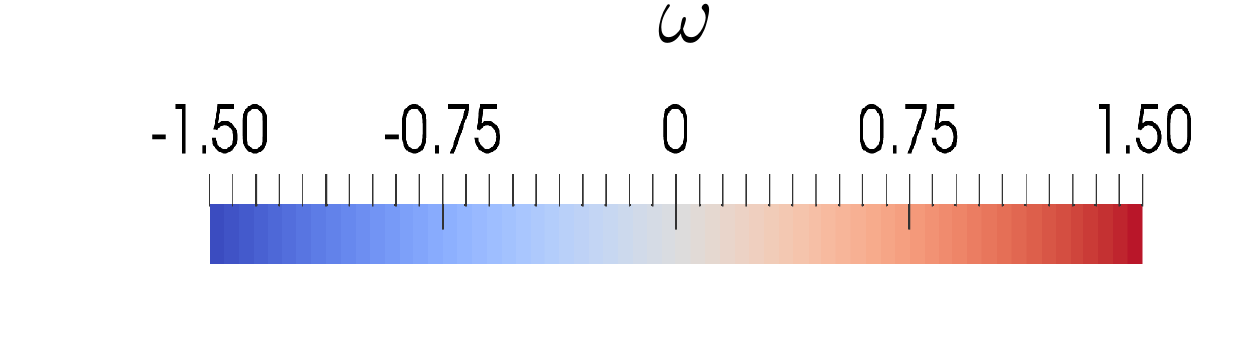} 
\hspace{0.3cm}
	\includegraphics[width=0.3\textwidth]{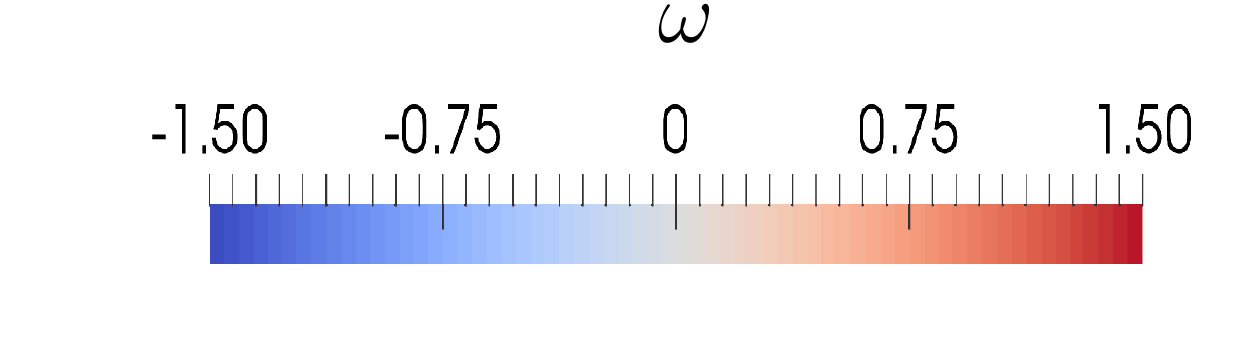} 
\hspace{0.3cm}
	\includegraphics[width=0.3\textwidth]{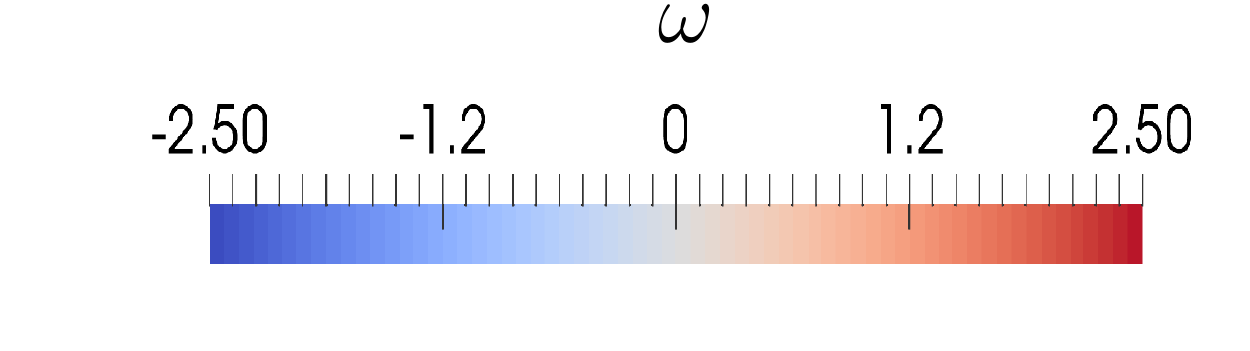} \\ 
	\includegraphics[width=0.3\textwidth]{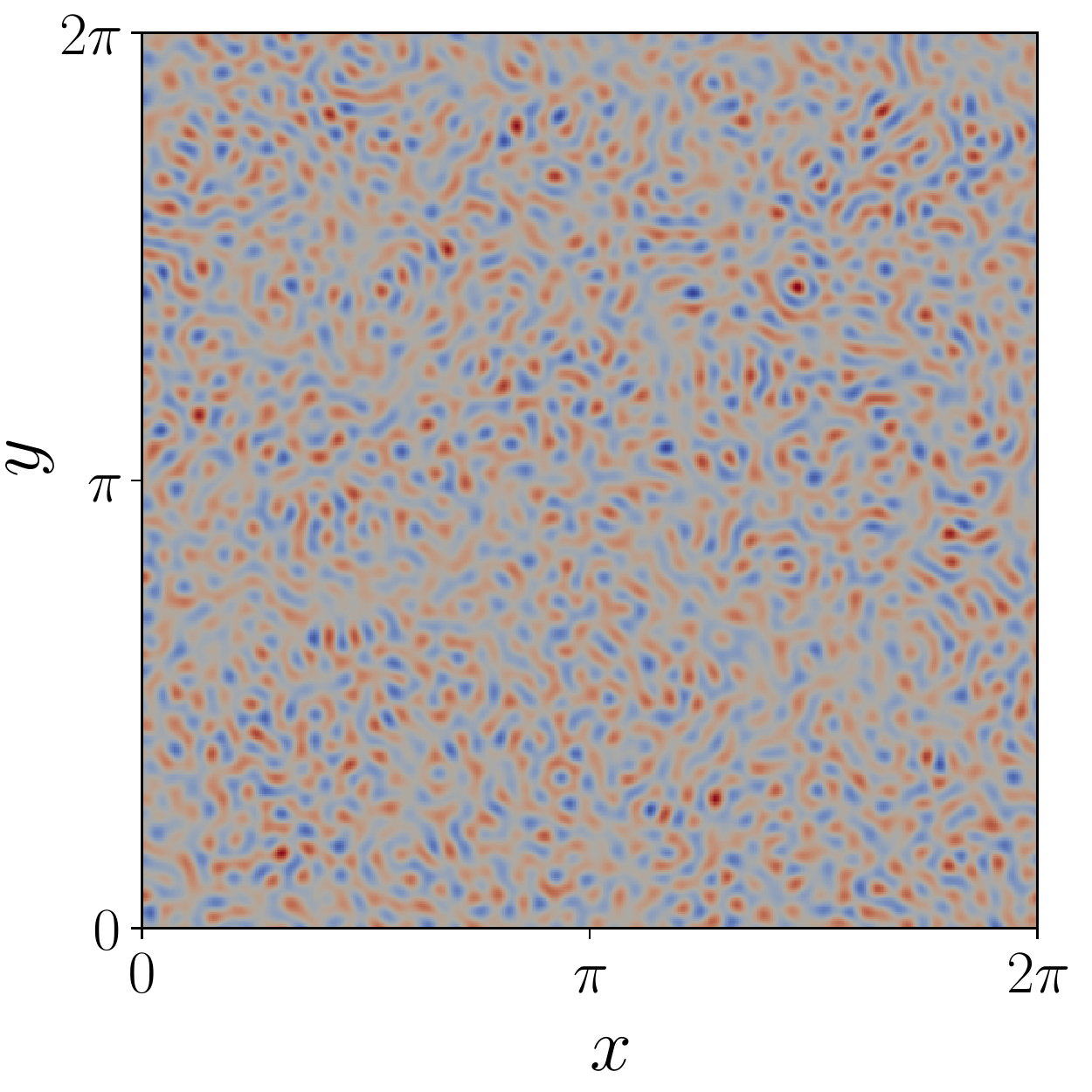} 
\hspace{0.3cm}
	\includegraphics[width=0.3\textwidth]{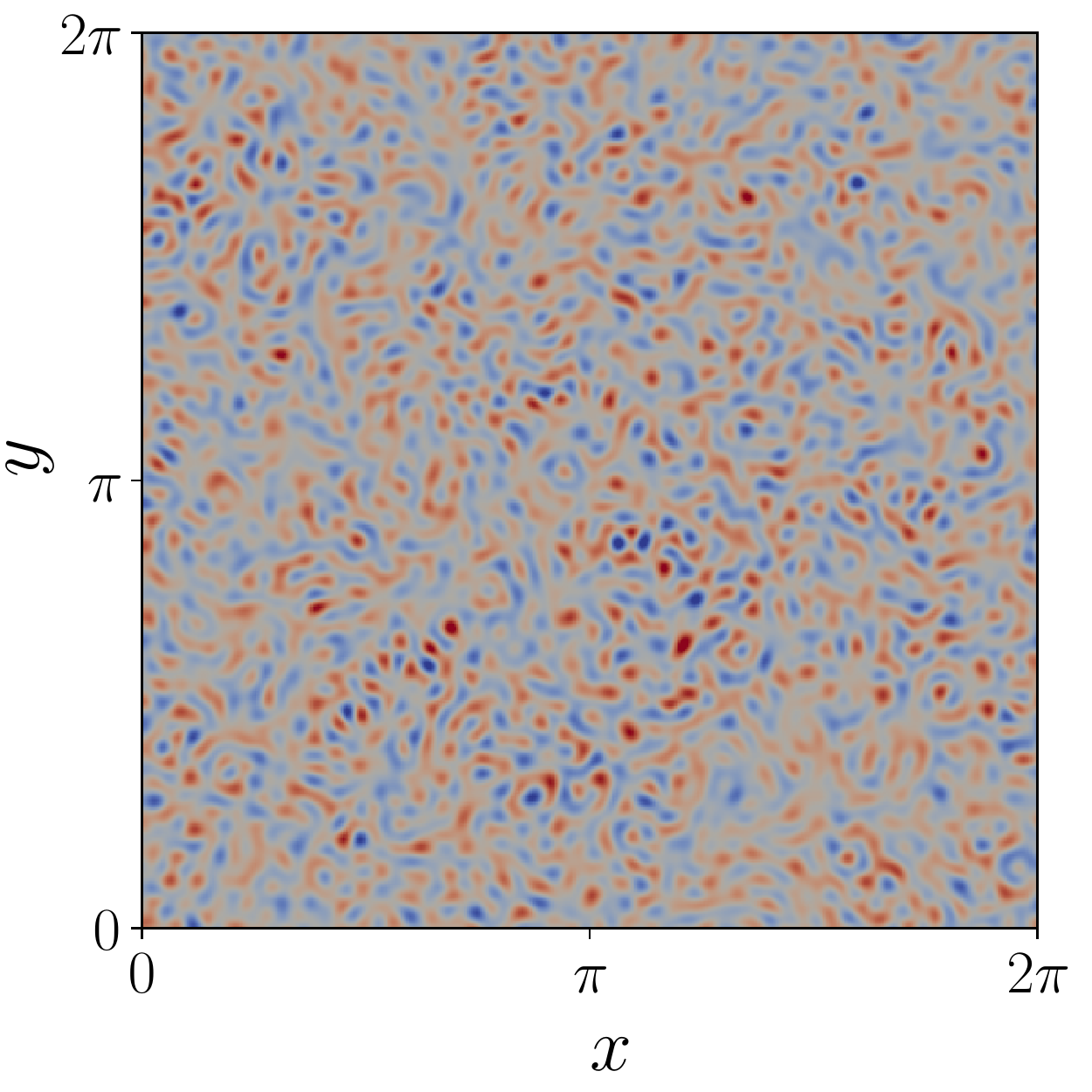} 
\hspace{0.3cm}
	\includegraphics[width=0.3\textwidth]{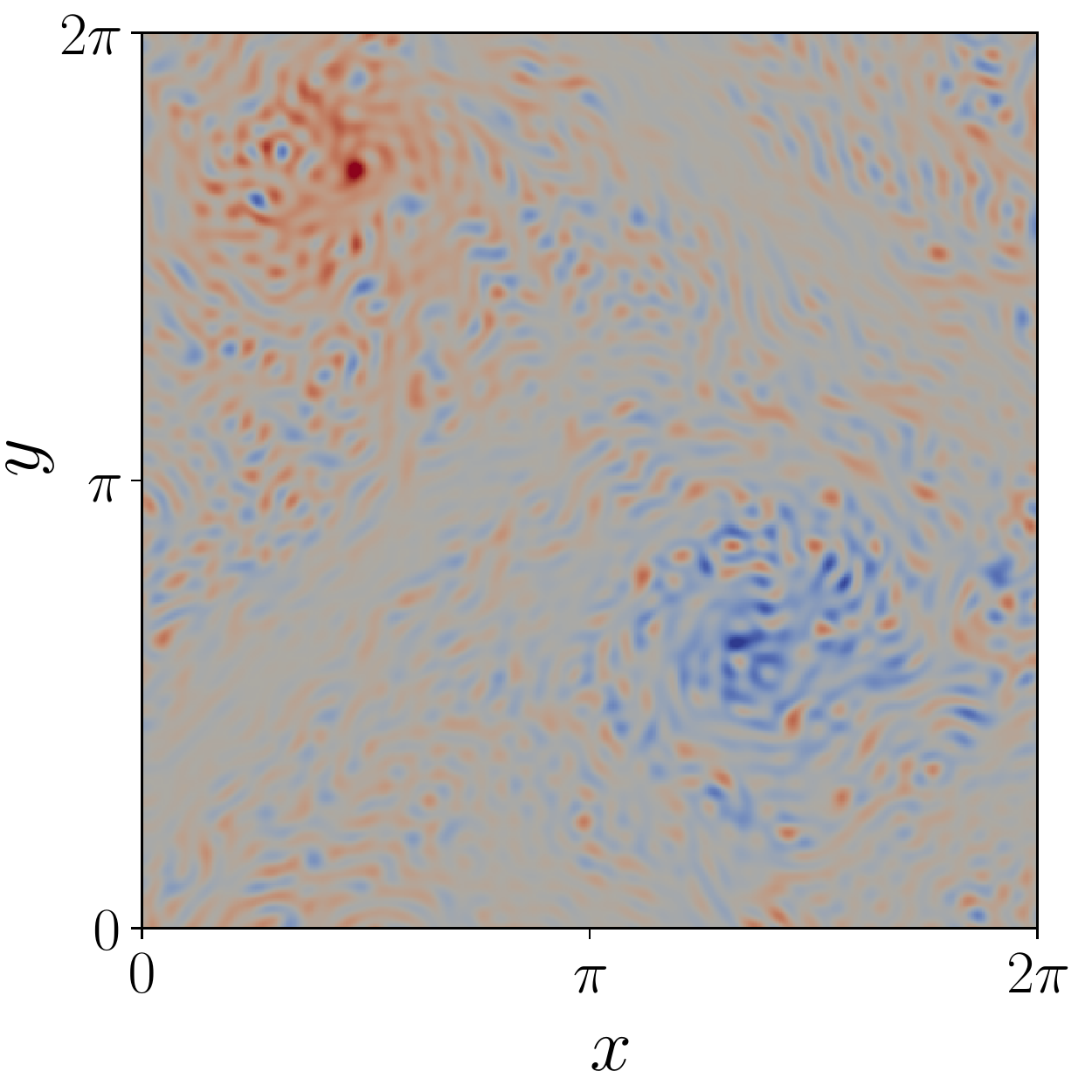} \\ 
\hspace{0cm} $\nu_1/\nu_0 = 1$ \hspace{4cm} $\nu_1/\nu_0 = 2$ \hspace{4cm} $\nu_1/\nu_0 = 5$ 

	\caption{(Color online) Visualisation of the vorticity field $\omega(x,y) \fvec{z}$ for PCV cases 
          $\nu_1/\nu_0 = 1$ , $\nu_1/\nu_0 = 2$ and $\nu_1/\nu_0 = 5$  
          (from left to right) using samples 
          taken during the statistically stationary state. 
	  }
 \label{fig:visuA0A2A14}
\end{figure*}

\subsection{Emergence of large-scale structures}
\label{sec:structure-formation}
The formation of successively larger structures 
and the eventual formation of a condensate 
with increasing amplification
can be seen in 
visualisations of the 
velocity field, 
as given in \cite{Linkmann19a}. Here, we provide 
visualisations of 
$\omega$ for the
three PCV cases in Fig.~\ref{fig:visuA0A2A14}.
The vorticity fields 
for $\nu_1/\nu_0 = 1$  and $\nu_1/\nu_0 = 2$ are similar, 
with the vortices in the latter case slightly stronger and a bit larger.
Finally, for
$\nu_1/\nu_0 = 5$ a condensate manifests itself in form of two counter-rotating
vortices as in classical 2d turbulence \cite{Smith93,Boffetta14ARFM}.

\begin{figure}[tbp]
\centering
	\includegraphics[width=\columnwidth]{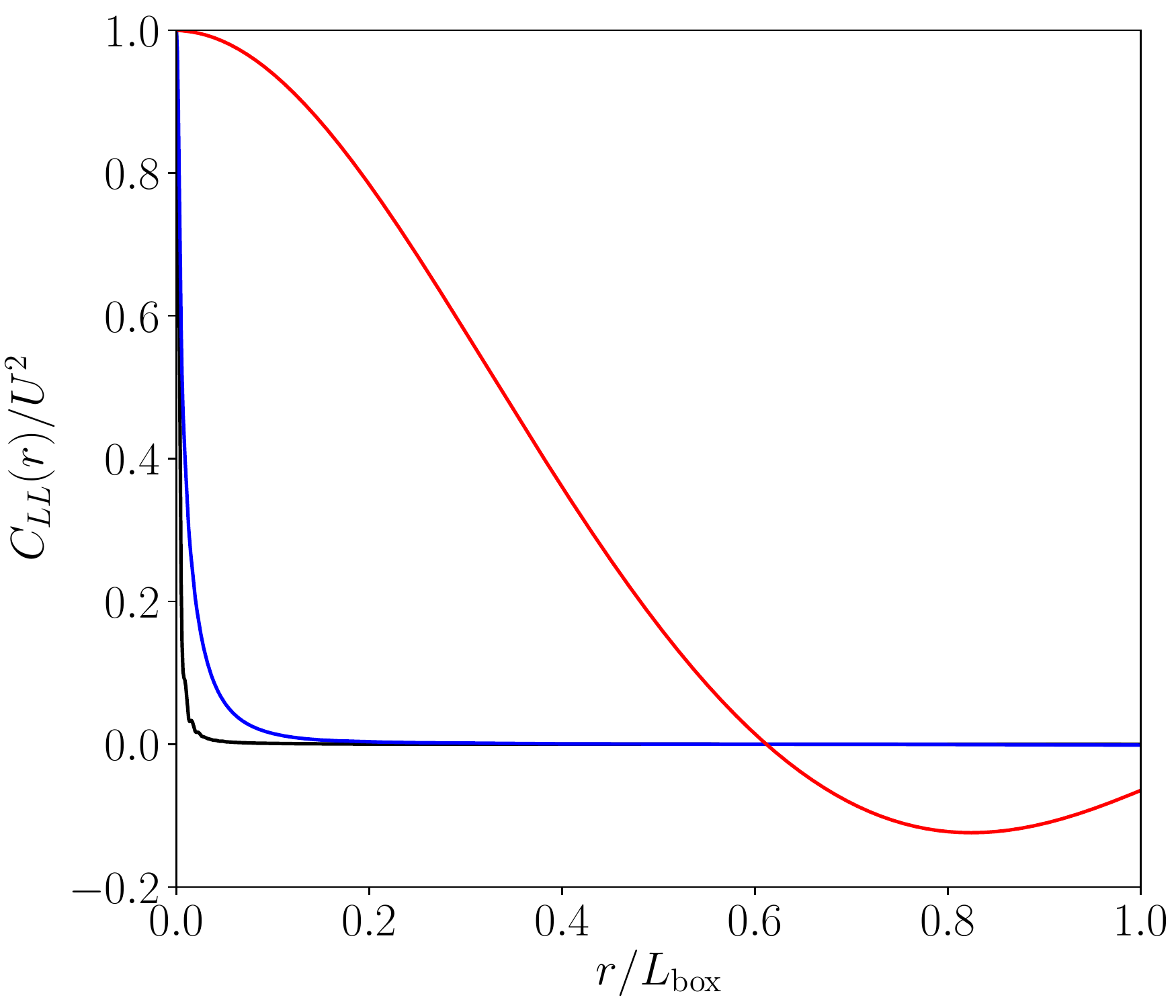} 
	\includegraphics[width=\columnwidth]{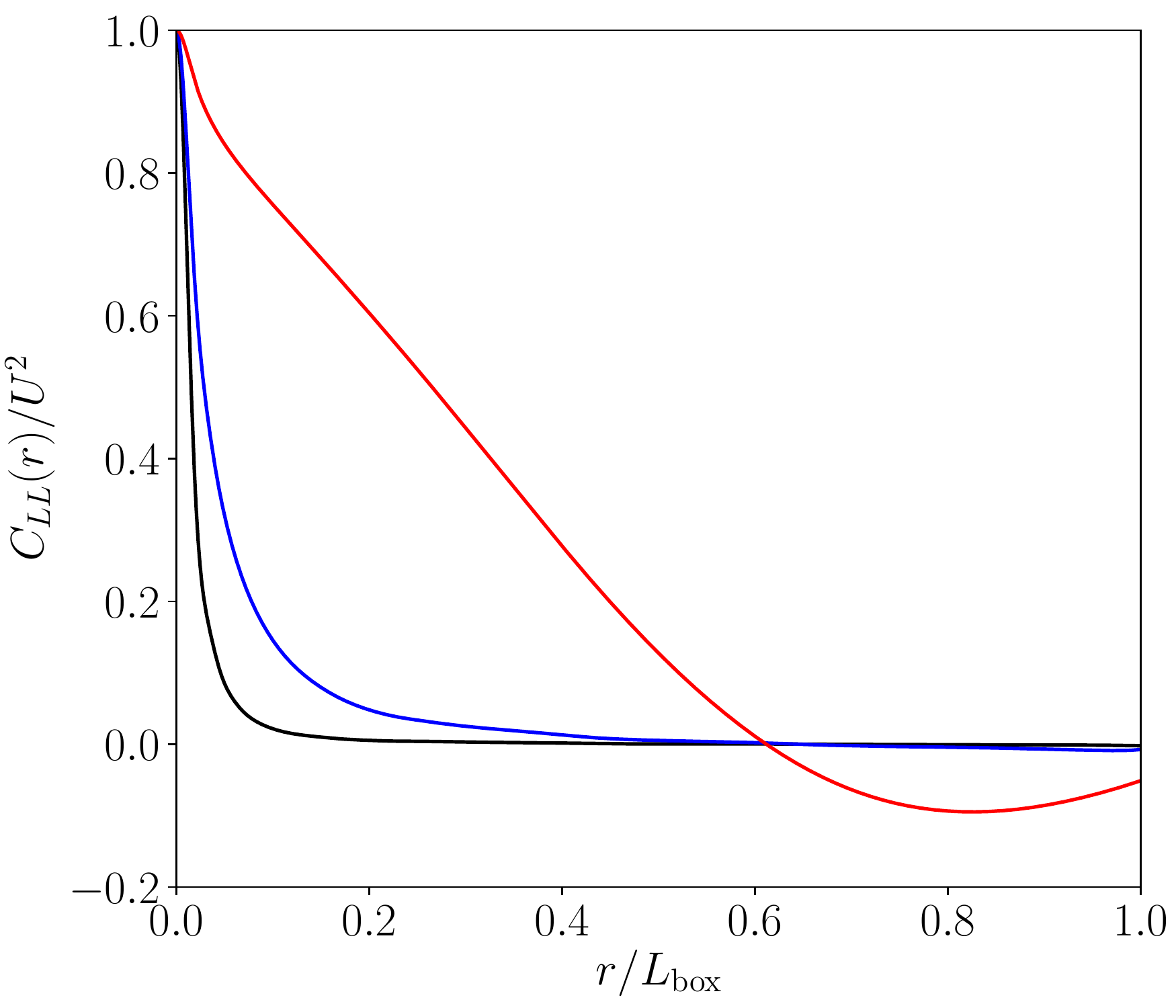} 
 \caption{
(Color online) Longitudinal correlation functions. Top: PCV for different values of $\nu_1/\nu_0$.
	Bottom: PEV for different values of $\Gamma_2/\Gamma_0$. 
         }
 \label{fig:correlations}
\end{figure}

The emergence of large-scale organization and coherence can be quantified 
through the calculation of equal-time correlation functions. Owing to 
isotropy, it is sufficient to consider the two-point longitudinal correlator
\beq
\gb{C_{LL}(r) = \langle u_L(\vec{x} + \vec{r}) u_L(\vec{x}) \rangle \ ,}
\eeq 
\gb{
where $r = |\vec{r}|$, and $u_L = \vec{u} \cdot \vec{r}/r$ is the velocity 
component along the displacement vector $\vec{r}$, } 
and the angled brackets denote a combined spatial and temporal average.  
Longitudinal correlation functions have been calculated through the spectral 
expansions of the respective 
velocity fields for PCV and PEV, with results shown in 
Fig.~\ref{fig:correlations}, where PCV and PEV data are contained in the 
top and bottom panels, respectively. 
Clear correlations up to the size of the system can be identified 
for $\nu_1/\nu_0 = 5$ and $\Gamma_2/\Gamma_0 = 0.0025$, 
while $C_{LL}$ decreases much faster in $r$ for 
the cases without a condensate, $\nu_1/\nu_0 = 1$, $\nu_1/\nu_0 = 2$,
$\Gamma_2/\Gamma_0 = 0.002$ and $\Gamma_2/\Gamma_0 = 0.0023$. 

The differences in correlation can also be quantified with
the integral scale
\beq
L \equiv \frac{1}{\gb{C_{LL}(0)}} \int_0^\infty dr \ C_{LL}(r)  \ , 
\eeq  
listed in table \ref{tab:simulations}: 
There is at least an $O(10)$ difference between the respective values of $L$ for 
PCV-B3 and the two cases with less amplification, PCV-B1 and PCV-B2, 
and similarly for PEV.
         
\section{Transition}
\label{sec:transition}
\begin{figure}[tbp]
\centering
	\includegraphics[width=\columnwidth]{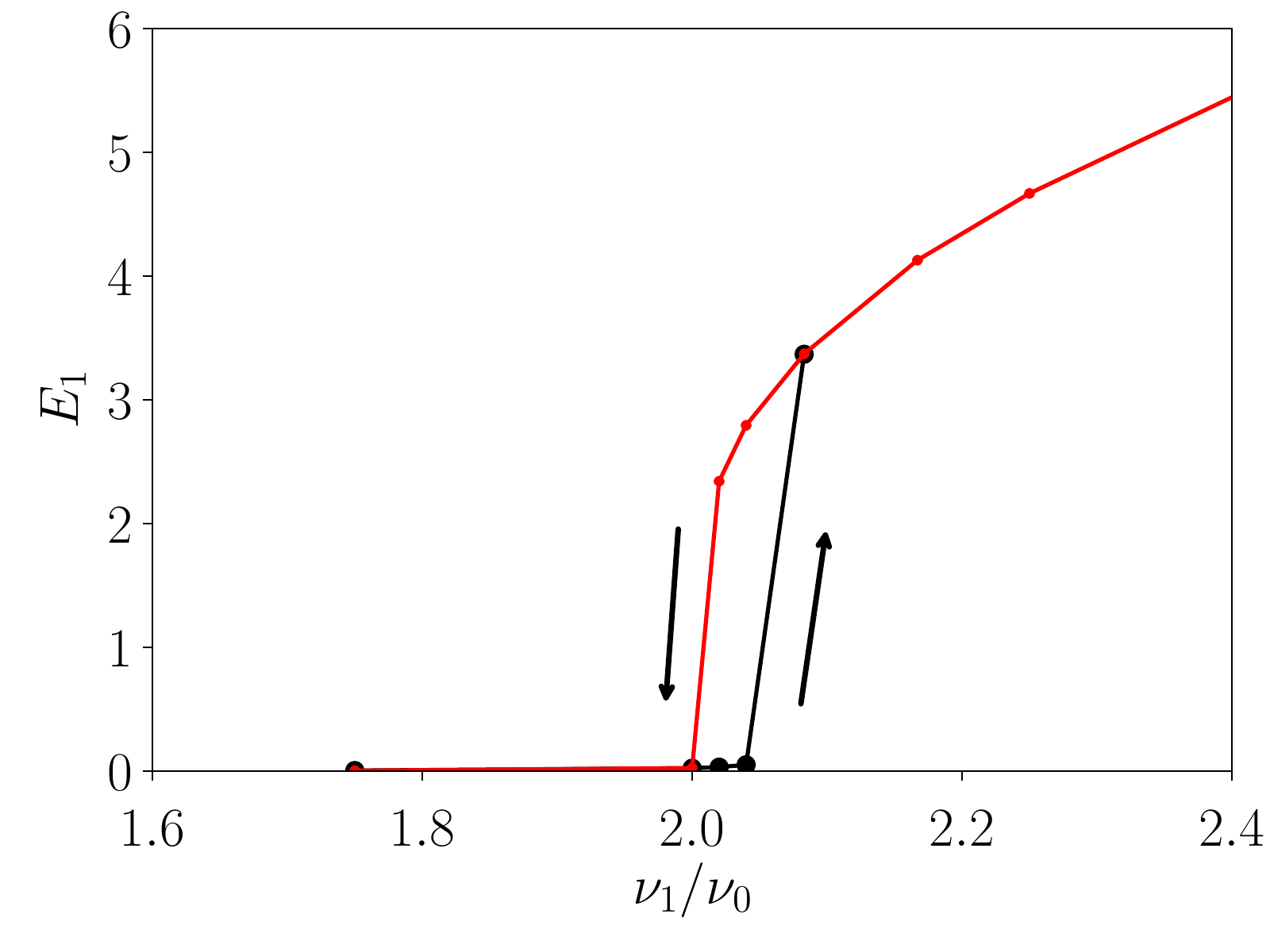} 
 \caption{
	 $E_1$ as a function of $\nu_1/\nu_0$. The black curve 
         corresponds to flow states obtained by increasing $\nu_1/\nu_0$ and 
	 the red (gray) curve to flow states obtained by decreasing $\nu_1/\nu_0$. 
         A hysteresis loop is visible in the region $2.00 \leqslant \nu_1/\nu_0 \leqslant 2.04$.
         }
 \label{fig:hysteresis}
\end{figure}

The transition between the two cases 
$\nu_1/\nu_0 = 1$ and $\nu_1/\nu_0 = 2$ 
without a condensate and $\nu_1/\nu_0=5$ 
with a condensate  
is discontinuous, as shown 
in \cite{Linkmann19a}, 

The 
discontinuous transition between spatiotemporal chaos and classical 2d-turbulence
suggests that the two states are separated by a subcritical bifurcation. 
Accordingly, 
we expect to find a bistable scenario with the possibility of coexisting
states in a  parameter range around the transition, and eventually also hysteresis.  
As observable we take the energy at the largest scale, $E_1$, which will be considered
as a function of the amplification factor and the energy input. $E_1$ is
calculated in terms of the energy spectrum
\beq
\label{eq:spectrum}
E(k) \equiv \left\langle
\frac{1}{2} \int d\hat{\vec{k}} \ |\fvec{u}(\vec{k})|^2
\right\rangle_t \ ,
\eeq
where $\int d\hat{\vec{k}}$ indicates an average over all angles in 
$k$-space with prescribed $|\vec{k}|=k$ and 
$\langle \cdot \rangle_t$ denotes a time average. $E_1$ is then given by 
$E_1 = E(k)_{|k=1}$.
Following our analysis in Ref.~\cite{Linkmann19a}, Fig.~\ref{fig:hysteresis}
presents $E_1$ as a function of $\nu_1/\nu_0$ close to the critical point. Two
main features of the transition can be identified in the figure.  First, $E_1$
increases suddenly at the critical value $\nu_1/\nu_0 = 2.00 \pm 0.02$, as
observed in Ref.~\cite{Linkmann19a}.  Second, the system shows hysteretic
behavior: The red (gray) curve consists of data points obtained for decreasing
$\nu_1/\nu_0$, while the black curve corresponds to states obtained for
increasing $\nu_1/\nu_0$. The resulting hysteresis loop is clearly visible. 

\begin{figure}[tbp]
\centering
	\includegraphics[width=\columnwidth]{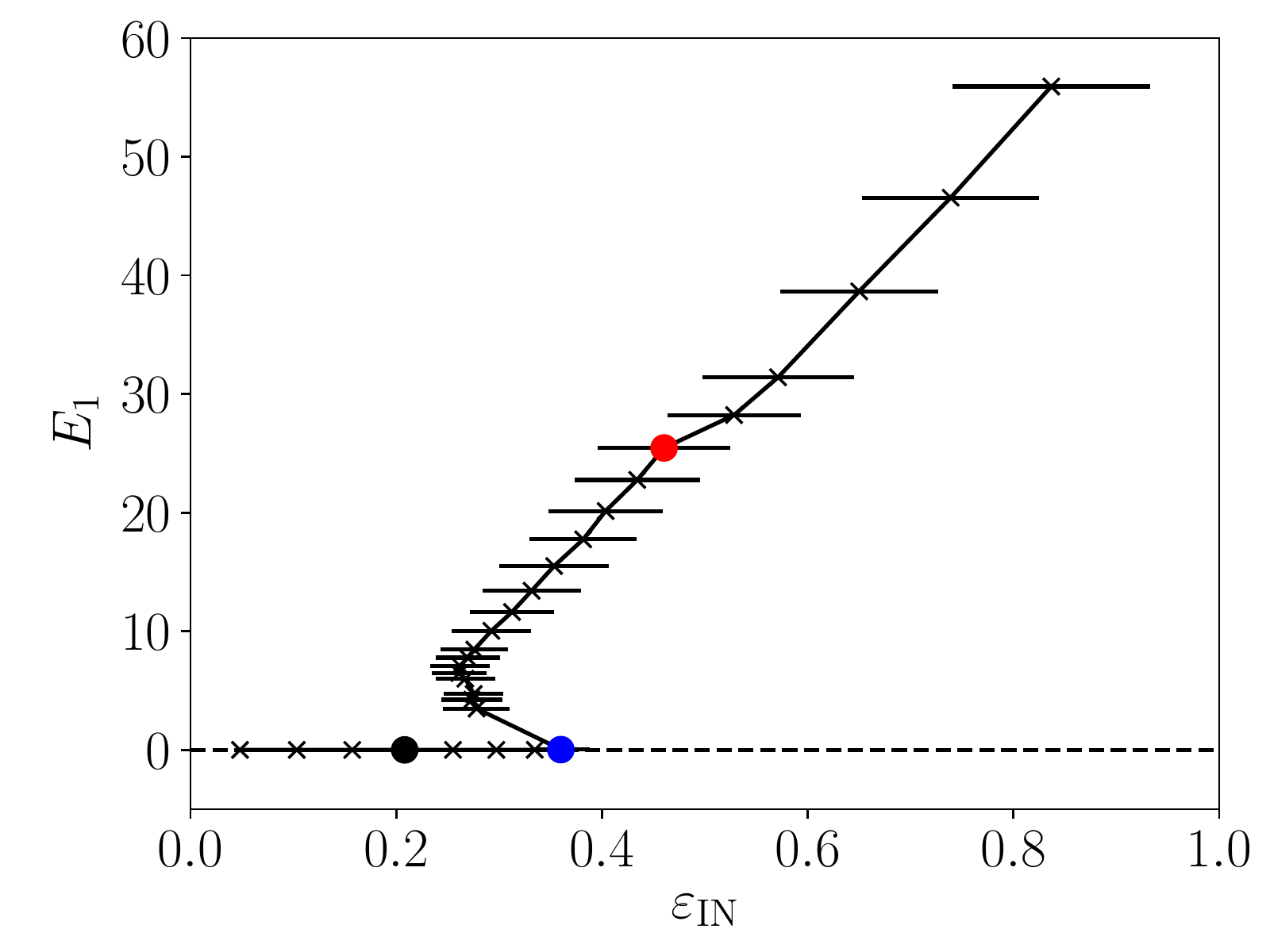} 
	\includegraphics[width=\columnwidth]{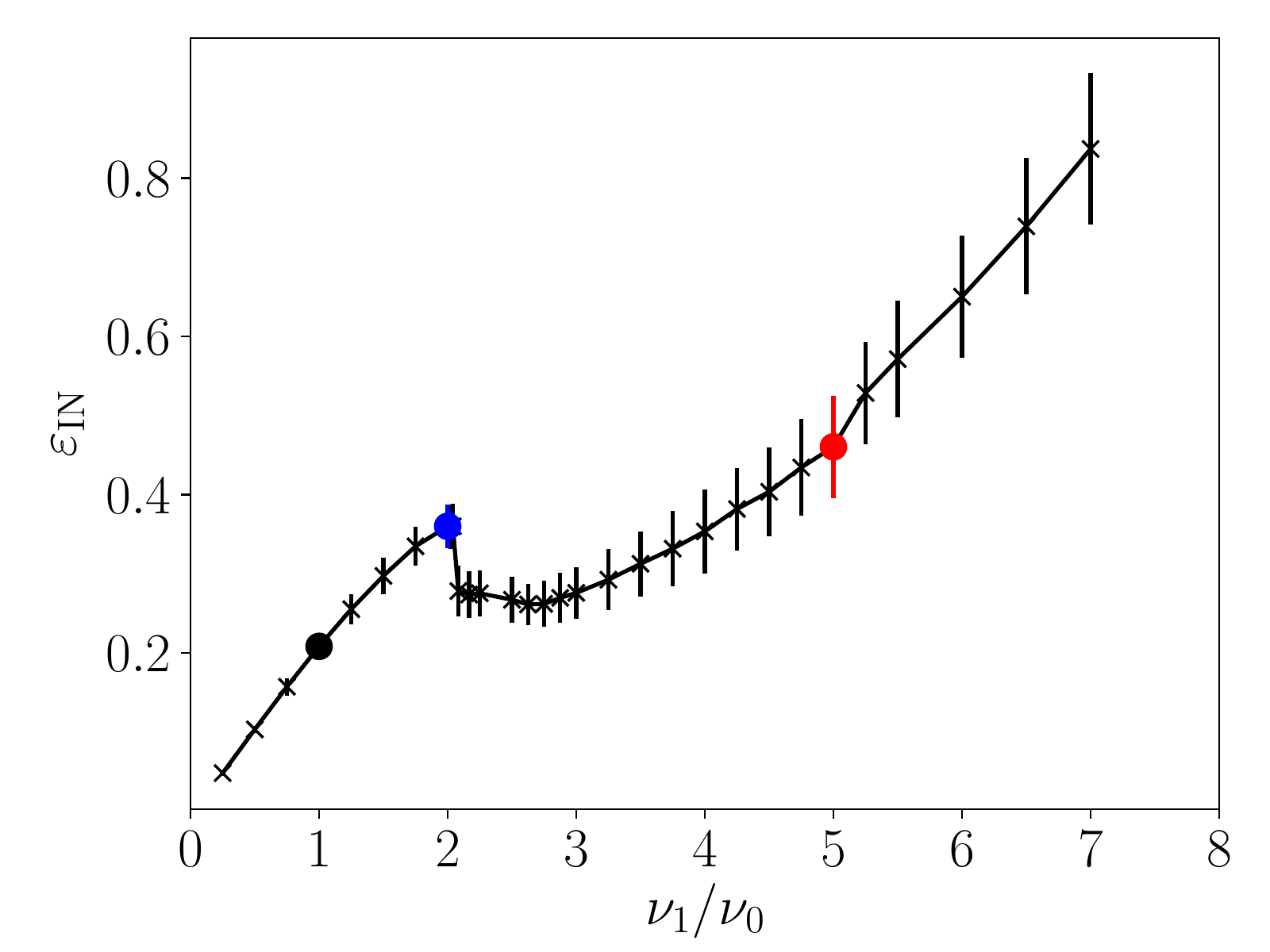} 
 \caption{
	 (Color online)
	 Top: $E_1$ as a function of $\varepsilon_{\rm IN}$.
         Bottom: $\varepsilon_{\rm IN}$ as a function of $\nu_1/\nu_0$.
	 The red (light gray), blue (dark gray) and black dots correspond to the  
	 PCV cases discussed in Sec.~\ref{sec:model-dynamics}.
 }
 \label{fig:E1_vs_epsIN}
\end{figure}

Apart from the presence of hysteresis shown here, the expected 
bistable scenario is realised in the statistically stationary total energy balance, 
\beq
\varepsilon = \varepsilon_{\rm IN} \simeq 2 \nu_0 \frac{(2\pi)^2}{L_f^2} \Ein \ ,  
\eeq
where
\beq
\Ein = \int_{k_{\rm min}}^{k_{\rm max}} dk \ E(k) \ ,
\eeq
with an upper and a lower branch of $\eps$ as a function of $\Ein$ 
corresponding to classical 2d turbulence with an emerging condensate and 
spatiotemporal chaos at the forcing scale, respectively, 
\cite{Linkmann19a}. 
The two branches were found to be connected by an unstable S-shaped region. 
The existence of two branches connected by an S-shaped region 
is also visible in the phase-space projection relating 
the energy at the largest scale to the energy input, i.e. 
for $E_1$ as a function of $\eps_{\rm IN}$ as shown 
in the top panel of Fig.~\ref{fig:E1_vs_epsIN}. The lower 
branch corresponds to injection rates obtained for 
$\nu_1/\nu_0 < \nu_{1,\rm crit}/\nu_0$, where $E_1$
is negligible and the inverse transfer is damped by dissipation 
at intermediate scales before reaching the largest scale in the
system. On the upper branch that describes states with a sizeable 
condensate, we observe a linear relation between 
$E_1$ and $\eps_{\rm IN}$, as can be expected if most energy is dissipated 
in the condensate
\beq
\epsin = \varepsilon \simeq 2 \nu_0 E_1 k_1^2 \Delta k \ ,  
\eeq
where $k_1 = 1$ is the lowest wavenumber in the domain, and $\Delta k = 1$ the 
width of the wavenumber shell centered at $k_1$.


The S-shaped region in the top panel of Fig.~\ref{fig:E1_vs_epsIN} can only
occur if $\eps_{\rm IN}$ is a non-monotonous function of the amplification
factor. This is indeed the case as can be seen in the bottom panel of the same
figure, where a sudden decrease in $\eps_{\rm IN}$ occurs at $\nu_{1,\rm
crit}/\nu_0$, followed by an interval in $\nu_1/\nu_0$ where $\eps_{\rm IN}$
varies very little. Eventually, for states with a strong condensate $\eps_{\rm
IN}$ increases linearly with $\nu_1/\nu_0$.  The nature of the transition is
thus related to non-monotonous behavior of the energy input (and therefore the
dissipation) as a function of the control parameter, which can only occur if
the energy input depends on the velocity field. In particular, for
Gaussian-distributed and $\delta$-in-time correlated forcing $\eps_{\rm IN}$
itself is the control parameter and a scenario as described here is unlikely to
occur.  This observation suggests that the type of transition depends on the
type of forcing, that is, it is non-universal. 

As explained in Sec.~\ref{sec:poly-model}, the structure of the 
PEV model make a parameter study with fixed energy input range difficult. 
However, the nature of the transition is unlikely
to be affected by the simplifications made in the PCV model, as 
the PEV and PCV models have the same structure in the sense that
energy input is given by linear amplification. 
The PEV simulations also show a sudden formation of a condensate 
under small changes in the amplification as can be seen from the 
comparison of correlation functions in the bottom panel of 
Fig.~\ref{fig:correlations}.

To compare to experimental data and between the two models, we define 
a Reynolds number based on the effective driving scale, $L_{\rm f}$ and 
the velocity at the driven scales 
\beq
{\rm Re}_f = \frac{\sqrt{\Ein}L_{\rm f}}{\tilde\nu} \ ,
\eeq
where $\tilde \nu$ is the Newtonian viscosity, i.e.  $\tilde \nu = \nu_0$ for
PCV and $\tilde \nu = \Gamma_0$ for PEV. This Reynolds number corresponds to
the Reynolds number associated with the mesoscale vortices observed in
experiments.  Values of ${\rm Re}_f$ for all simulations are given in table
\ref{tab:simulations}. The transition occurs at ${\rm Re}_f \simeq 20$ for PCV
and at ${\rm Re}_f \simeq 10$ for PEV, the exact value may depend on simulation
details such at the width of the driving range and the level of small-scale
dissipation. However, the main point is that both models transition at Reynolds
number of $O(10)$. In comparison, the experimentally observed Reynolds numbers
are about $O(10^{-2})$, based on characteristic vortex sizes of $100 \mu m$,
with a characteristic speed of $100\mu m/s$ for {\em B. subtilis}
\cite{Dombrowski04}, and the kinematic viscosity of water $\nu_{\rm
H_2O}=10^{-6} (\mu m)^2/s$.

\subsection{Spectral scaling}
\label{sec:spectral-scaling}

\begin{figure}[tbp]
\centering
	\includegraphics[width=\columnwidth]{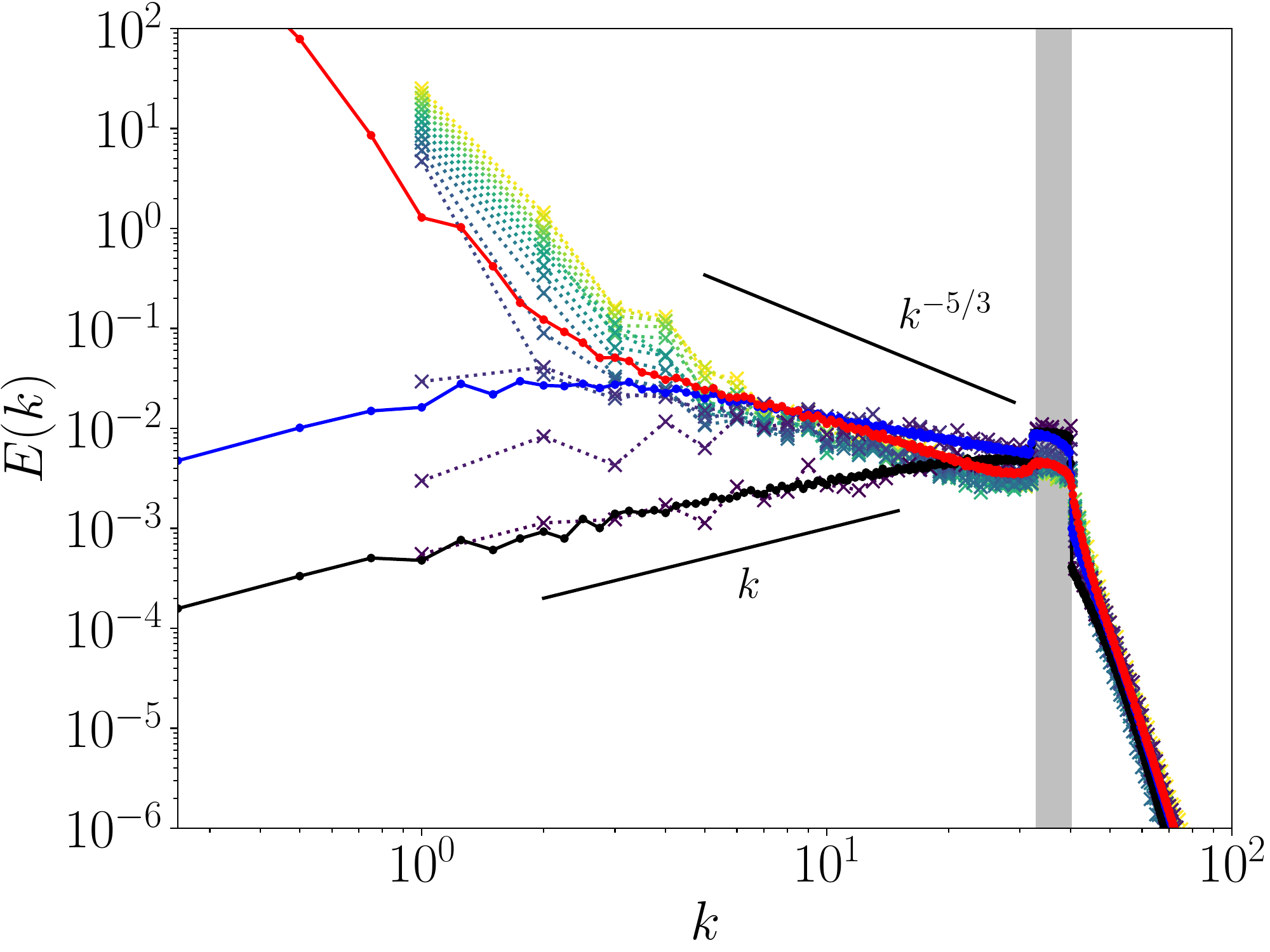} 
	\includegraphics[width=\columnwidth]{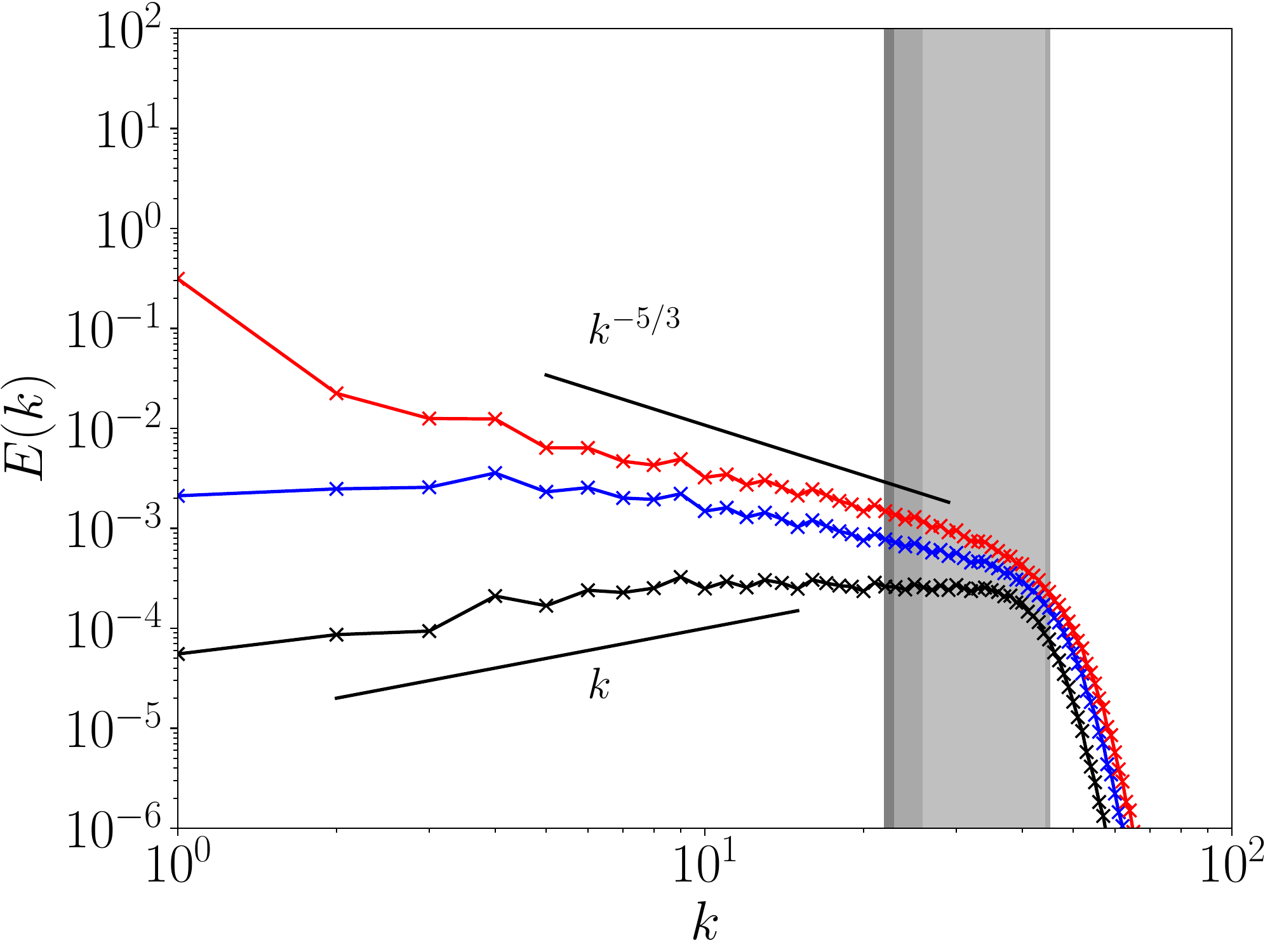} 
 \caption{
	 (Color online) Energy spectra. Top: PCV with for different values of 
	 $\nu_1/\nu_0$. 
	 The solid lines show the rescaled PCV-B cases 
	 $\nu_1/\nu_0 = 1$ (black) , $\nu_1/\nu_0 = 2$ (blue) 
	 and $\nu_1/\nu_0 = 5$ (red), and the dotted lines PCV-A data. 
	 Bottom: PEV with $\Gamma_2/\Gamma_0 = 0.0025$ (red), $\Gamma_2/\Gamma_0 = 0.0023$ (blue) and
         $\Gamma_2/\Gamma_0 = 0.002$ (black).
	 The grey-shaded areas indicate the respective driving ranges.
         }
 \label{fig:espectra}
\end{figure}

Energy spectra for PCV and PEV are shown in the top and bottom panels of 
Fig.~\ref{fig:espectra}, respectively. The dotted lines in the top panel 
correspond to series PCV-A, and the solid lines to rescaled PCV-B data as in Ref~\cite{Linkmann19a}.
The transition can be located clearly in the spectra as $E_1$ increases by three orders of 
magnitude from the third to the fourth dotted line.  
The PEV energy spectra in the bottom panel \mcm{of Fig.~\ref{fig:espectra}} 
correspond to $\Gamma_2/\Gamma_0 = 0.0025$ (red),
$\Gamma_2/\Gamma_0 = 0.0023$ (blue) and $\Gamma_2/\Gamma_0 = 0.002$ (black),
with the forcing centered around $k_{\rm f}=36$ as in the PCV model. The
results are similar to those for the PCV model shown in the top panel of
Fig.~\ref{fig:espectra}: A condensate forms suddenly under small changes
in the amplification.  This further corroborates that the existence and the
nature of the transition do not depend on the simplifications of the PEV model
that led to the construction of the PCV model.  Energy spectra with an extended
scaling range and a small accumulation of energy at the smallest wave number
have also been observed in the bacterial flow model \cite{Oza2016}.  There, the
critical amplification rate at which the condensate occurs will depend on the
\mcm{relaxation} 
term $-\alpha_F \vp$ that originates from the functional derivative of
the free energy given in Eq.~\eqref{eq:free-energy}. Indeed, the existence of a
critical value of $\alpha_F > 0$, below which no energy accumulation occurs, has
been reported in Ref.~\cite{Bratanov15}. Similarly, condensate formation in Newtonian 
turbulence can be suppressed in presence of sufficiently strong linear friction \cite{Danilov01}.
In view of the transition scenarios, a general quantification of the effect of 
large-scale dissipation would be of interest. 

At low amplification, equipartition scaling $E(k) \propto k$ is observed for
PEV and PCV, as indicated by the black curves in Fig.~\ref{fig:espectra}. In
contrast, the low-wavenumber form of $E(k)$ is non-universal for the bacterial
flow model even at very low amplification \cite{Bratanov15}.  This difference
also originates from the presence of the \mcm{relaxation} 
term  $-\alpha_F \vp$  in the
bacterial flow model, in Ref.~\cite{Bratanov15} the scaling exponent of $E(k)$
at $k < k_{\rm min}$ is found to depend on $\alpha_F$. 
In Newtonian turbulence, deviations from Kolmogorov-scaling of $E(k)$ also depend
on details of large-scale dissipation such as the strength of a linear friction term 
or the use of hypoviscosity \cite{Danilov01}.

Further observations can be made from the data shown in Fig.~\ref{fig:espectra}.
The spectral exponent is larger than the Kolmogorov value of $-5/3$ 
even in presence of an inverse energy transfer, resulting in shallower spectra. 
This can have several reasons. For simulations with a small condensate such as for the PEV dataset with 
$\Gamma_2/\Gamma_0 = 0.0025$ shown in red (light gray) in the bottom panel, 
energy dissipation is not negligible in the wavenumber range between the condensate and the 
driven interval, and Kolmogorov's hypotheses do not apply. For simulations
with a sizeable condensate such as PCV-B3 shown in red (light gray) in the top 
panel, the condensate itself alters the dynamics in the inertial range. In presence of a 
strong condensate the spectral scaling is known to become steeper \cite{Chertkov07}, 
with $E(k) \propto k^{-3}$ for the entire wavenumber range $k < k_{\rm min}$. 
Removing the coherent part of the velocity field results in shallower scaling $E(k) \propto k^{-1}$ \cite{Chertkov07}.  
Intermediate states with spectra similar to PCV-B3 have also been obtained, see Fig.~3A 
in Ref.~\cite{Chertkov07}.

\subsection{Nonlocal transfers}
\label{sec:nonlocal}
\begin{figure*}[tbp]
\centering
	\includegraphics[width=0.45\textwidth]{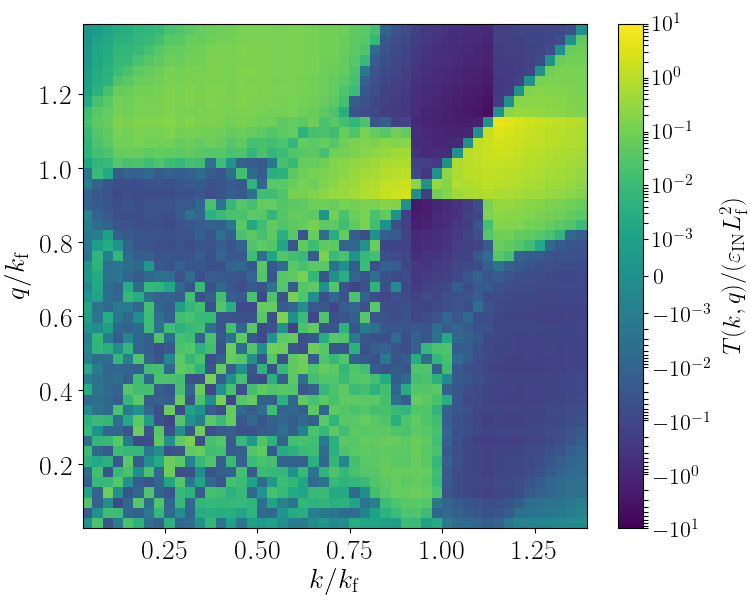} 
	\includegraphics[width=0.45\textwidth]{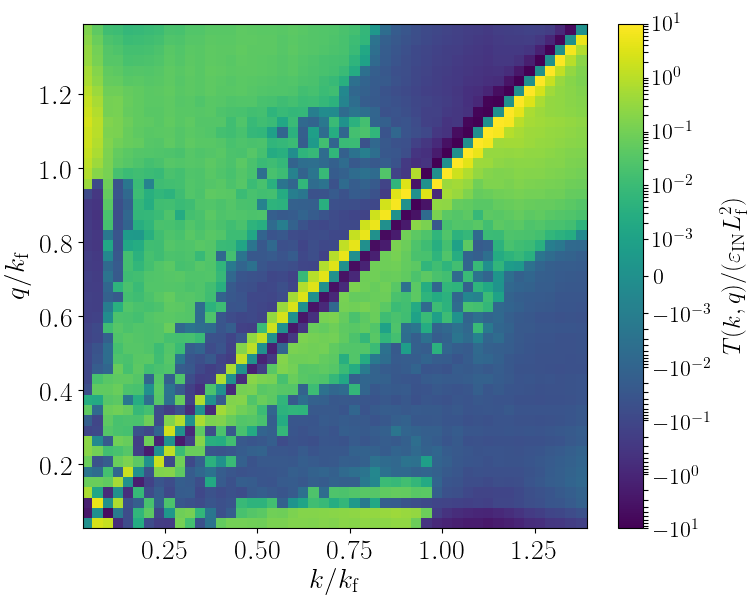} 
	\caption{
         (Color online) PCV shell-to-shell transfer function $T(k,q)/(\epsin L_{\rm f}^2)$ for $1 \leqslant k,q \leqslant 50$. 
	 Left: $\nu_1/\nu_0 = 2$, without condensate. Right: $\nu_1/\nu_0 = 5$, with condensate at $k=1$.
         }
 \label{fig:E13-transfer_kq}
\end{figure*}

Since the driving in both models depends on the amount of energy in the driven
range, a reduction in the energy input with {\em increasing} amplification
requires a reduction in $\Ein$. One way by which this could happen is through
an enhanced nonlinear transfer out of the driven wave number range. The
reduction in $\Ein$ occurs at the critical point, which suggests that the
condensate may couple directly to the driven scales, leading to a non-local
spectral energy transfer from the driven wave number interval into the
condensate.  In order to investigate whether this is the case, the energy
transfer spectrum 
was decomposed into shell-to-shell transfers \cite{Domaradzki90,Bratanov15} between 
linearly spaced spherical shells centered at wavenumbers $k$ and $q$
\beq
T(k,q) = \int d \fvec{k} \int d \fvec{q} \int d \vec{p} \ 
\fvec{u}_{\vec{k}}^* \cdot (\fvec{u}_{\vec{p}}\cdot i\vec{q})\fvec{u}_{\vec{q}}\delta(\vec{k} + \vec{p} - \vec{q}) \ , 
\eeq
\gb{where $\fvec{k}$ and $\fvec{q}$ are unit vectors.}
Here, the focus is on the existence of a coupling between the condensate and
the driven scales, hence linear shell-spacing is sufficient. More quantitative
statements concerning the relative weight of different couplings within the
overall transfer requires logarithmic spacing \cite{Aluie09}.  Figure
\ref{fig:E13-transfer_kq} shows the non-dimensional transfer $T(k,q)/(\epsin
L_{\rm f}^2)$ for two example cases, one without condensate (left panel) and one
with condensate (right panel). In both cases the transfers are antisymmetric
about the diagonal. This must be the case, as energy conservation requires
$T(k,q)$ to be antisymmetric under the exchange of $k$ and $q$. As can be seen
in the left panel of Fig.~\ref{fig:E13-transfer_kq}, in absence of a condensate
$T(k,q)$ is concentrated along the diagonal, that is energy is mainly
redistributed locally and close to the driven scales. In contrast, the transfers 
shown in the bottom panel of Fig.~\ref{fig:E13-transfer_kq} include off-diagonal 
contributions where the condensate couples directly to the driven wavenumber range. 

\section{Four-scale model}
\label{sec:four-scale-model}
Some of the qualitative features of the transition can be captured in a
four-scale model.  Let $\Els$ be the energy content at the intermediate
wavenumbers $1 < k < k_{\rm min})$  and $\Ess$ the energy content at $k >
k_{\rm max}$.  Then one can consider the interaction of the four quantities
$E_1, \Ein, \Els$ and $\Ess$  
\begin{align}
\label{eq:evol_econd}
\dot E_1  = & -2\nu_0 k_1^2 E_1  +c_3E_1^{1/2}\Els \nonumber \\  
            & +c_2\theta(E_1 - E_{1,0})(E_1 - E_{1,0})^{1/2}\Ein \ , \\ 
\label{eq:evol_els}
\dot \Els = & -2\nu_0 k_{\rm LS}^2\Els +c_1 \Els^{1/2}\Ein  -c_3E_1^{1/2}\Els \ , \\
\label{eq:evol_ein}
\dot \Ein = & 2\nu_1 k_{\rm IN}^2\Ein  -c_1 \Els^{1/2}\Ein -c_4 E_{\rm SS}^{1/2}E_{\rm IN} \nonumber \\  
            & -c_2\theta(E_1 - E_{1,0})(E_1 - E_{1,0})^{1/2}\Ein \ , \\ 
\label{eq:evol_ess}
\dot E_{\rm SS} = & -2\nu_2 k_{\rm SS}^2 E_{\rm SS} + c_4 E_{\rm SS}^{1/2}E_{\rm IN} \ , 
\end{align}
where $\theta$ is the Heaviside step function, $c_i > 0$ for $i = 1,\hdots,4$
parametrise the coupling terms and $k_1=1$, $k_{\rm LS}$, $k_{\rm IN}$ and $k_{\rm SS}$ are effective
wavenumbers in the corresponding ranges. In terms of energy transfers, the coupling terms represent
\begin{align}
& \Ein \longrightarrow \Els:  \quad c_1 \Els^{1/2}\Ein \ , \\ 
& \Ein \longrightarrow E_1:   \quad c_2 \theta(E_1 - E_{1,0})(E_1 - E_{1,0})^{1/2}\Ein \ , \\ 
& \Els \longrightarrow E_1:   \quad c_3 E_1^{1/2}\Els \ , \\
& \Ein \longrightarrow \Ess:  \quad c_4 \Ess^{1/2}\Ein \ , 
\end{align}
where the coupling between $\Ein$ and $E_1$ is modelled such that a nonlocal
energy transfer from the driven wavenumber range into the largest resolved
scales only takes place once a condensate is emerging.  The coupling parameters
$c_i$ can be obtained from DNS data through calculations of shell-to-shell
nonlinear transfers. Once they are known, a parameter scan in $\nu_1$ can be
carried out for different values of the threshold energy $E_{1,0}$ in order to
compare the results from the model with the DNS data.  However, before doing
so, we derive predictions from the model equations for two asymptotic cases:
\begin{itemize}
\item[(i)]  presence of a strong condensate, $E_1 \gg E_{1,0}$, corresponding to the upper
     branch in Fig.~\ref{fig:E1_vs_epsIN},
\item[(ii)] absence of a condensate $E_1 < E_{1,0}$, corresponding to the lower
     branch in Fig.~\ref{fig:E1_vs_epsIN}.
\end{itemize}
In what follows the small-scale dissipation is neglected, as this 
enables us to focus on the main points. We will come back to an analysis
of the full model in Sec.~\ref{sec:model-scan}.

\subsubsection{case {\rm (i)}: $E_1 \gg E_{1,0}$} 
\label{sec:fourscale-above}
For  $E_1 \gg E_{1,0}$ we approximate the coupling term between $\Ein$ and $E_1$ as
\beq
c_2 \theta(E_1 - E_{1,0})(E_1 - E_{1,0})^{1/2}\Ein \simeq c_2 E_1^{1/2}\Ein \ ,
\eeq
and we neglect the coupling term $c_3E_1^{1/2}\Els$ that describes a local energy transfer
from the intermediate scales into the condensate. 
The latter is introduced to model the nonlocal contribution to the
inverse energy transfer in presence of a condensate
as discussed in Sec.~\ref{sec:nonlocal}.
Equations \eqref{eq:evol_econd}-\eqref{eq:evol_ein} then simplify to
\begin{align}
\label{eq:evol_ein_upper}
\dot \Ein &= 2\nu_1 k_{\rm IN}^2\Ein  -c_1 \Els^{1/2}\Ein  -c_2 E_1^{1/2}\Ein \ , \\
\label{eq:evol_els_upper}
\dot \Els &= -2\nu_0 k_{\rm LS}^2\Els +c_1 \Els^{1/2}\Ein \ , \\
\label{eq:evol_econd_upper}
\dot E_1  &= -2\nu_0 k_1^2 E_1  +c_2 E_1^{1/2}\Ein \ ,
\end{align}
which result in the following expressions for $\Ein$, $\Els$ and $E_1$ in steady state
\begin{align}
\label{eq:SS_ein_upper}
	2\nu_1 k_{\rm IN}^2 \Ein & = c_1 \Els^{1/2}\Ein  +c_2 E_1^{1/2}\Ein \nonumber \\ 
	& \implies  c_1 \Els^{1/2}  +c_2 E_1^{1/2} = -2\nu_1k_{\rm IN}^2 \ , \\
\label{eq:SS_els_upper}
	2\nu_0 k_{\rm LS}^2\Els & = c_1 \Els^{1/2}\Ein \nonumber \\ 
	& \implies \Els = \left( \frac{c_1}{2\nu_0k_{\rm LS}^2} \Ein\right)^2 \ , \\
\label{eq:SS_econd_upper}
	2\nu_0 k_1^2E_1  & =  c_2 E_1^{1/2}\Ein \nonumber \\
	& \implies E_1 = \left( \frac{c_2}{2\nu_0k_1^2} \Ein\right)^2 \ .
\end{align}
Solving for $\Ein$ as a function of $\nu_1$, one obtains
\begin{align}
\label{eq:Einput_nu1_upper}
	2\nu_1 k_{\rm IN}^2 & = c_1 \Els^{1/2}  +c_2 E_1^{1/2} = \frac{\frac{c_1^2}{2k_{\rm LS}^2} + \frac{c_2^2}{2k_1^2}}{\nu_0}\Ein 
	\nonumber \\ 
	&\implies \Ein = -4\frac{\nu_1\nu_0 k_{\rm IN}^2}{\frac{c_1^2}{k_{\rm LS}^2} + \frac{c_2^2}{k_1^2}} \ ,
\end{align}
that is, $\Ein \sim \nu_1$ and $E_1 \sim \nu_1^2$, in qualitative agreement with the data presented in
Fig.~3 of Ref.~\cite{Linkmann19a} for $\nu_1 > \nu_{1,\rm crit}$, respectively.
\\

\subsubsection{case {\rm (ii)}: $E_1 < E_{1,0}$} 
\label{sec:fourscale-below}
In this case, there is no nonlocal coupling between $E_1$ and $\Ein$, hence
Eqs.~\eqref{eq:evol_econd}-\eqref{eq:evol_ein} become
\begin{align}
\label{eq:evol_ein_lower}
\dot \Ein &= 2\nu_1 k_{\rm IN}^2 \Ein  -c_1 \Els^{1/2}\Ein  \ , \\
\label{eq:evol_els_lower}
\dot \Els &= -2\nu_0 k_{\rm LS}^2 \Els +c_1 \Els^{1/2}\Ein  -c_3E_1^{1/2}\Els \ , \\
\label{eq:evol_econd_lower}
\dot E_1  &= -2\nu_0 k_1^2 E_1  +c_3E_1^{1/2}\Els  \ ,
\end{align}
which leads the the following expressions in steady state
\begin{align}
\label{eq:SS_evol_ein_lower}
	2\nu_1 k_{\rm IN}^2 \Ein & = c_1 \Els^{1/2}\Ein \nonumber \\ 
	& \implies \Els = \left ( \frac{2\nu_1 k_{\rm IN}^2}{c_1}\right)^2 \ , \\
\label{eq:SS_evol_els_lower}
	2\nu_0 k_{\rm LS}^2 \Els & = c_1 \Els^{1/2}\Ein  -c_3E_1^{1/2}\Els \nonumber \\ 
	& \implies \Ein = \frac{1}{c_1}\left(2\nu_0 k_{\rm LS}^2 + c_3E_1^{1/2} \right) \Els^{1/2}\ , \\
\label{eq:SS_evol_econd_lower}
	2\nu_0 k_1^2 E_1   & = c_3E_1^{1/2}\Els  \nonumber \\ 
	& \implies E_1 = \left (\frac{c_3}{2\nu_0 k_1^2} \Els \right)^2  \ .
\end{align}
Solving for $\Ein$ as a function of $\nu_1$, one obtains
\beq
\label{eq:Einput_nu1_lower}
\Ein = 4\frac{\nu_1\nu_0}{c_1^2}k_{\rm IN}^2\left(k_{\rm LS}^2+\left( \frac{c_3\nu_1k_{\rm IN}^2}{c_1\nu_0 k_1} \right)^2 \right) \ ,
\eeq
while $E_1 \sim \nu_1^4$. \\

\noindent
Comparing the energy content in the driven wavenumber range between cases (i) and (ii) given
in Eqs.~\eqref{eq:Einput_nu1_upper} and \eqref{eq:Einput_nu1_lower}, respectively,  we find
\begin{align}
\label{eq:model-drop}
	E_{\rm IN}^{\varepsilon_+} &= 4\frac{\nu_1\nu_0 k_{\rm IN}^2}{\frac{c_1^2}{k_{\rm LS}^2} + \frac{c_2^2}{k_1^2}}
	\nonumber \\
	& < 4\frac{\nu_1\nu_0}{c_1^2}k_{\rm IN}^2\left(k_{\rm LS}^2+\left( \frac{c_3\nu_1k_{\rm IN}^2}{c_1\nu_0 k_1} \right)^2 \right)
        = E_{\rm IN}^{\varepsilon_-}  \ .
\end{align}
We point out that this comparison is only justified close to the critical
point, as in principle the different cases imply different ranges of $\nu_1$:
case (i) is applicable for $\nu_1 > \nu_{1,\rm crit}$ and case (ii) for $\nu_1
< \nu_{1,\rm crit}$.  However, in the vicinity of $\nu_{1,\rm crit}$,
Eq.~\eqref{eq:model-drop} predicts a sudden drop in $\Ein$ and therefore of
$\eps_{\rm IN} = 2 \nu_1 k_{\rm IN}^2 \Ein$ as a function of $\nu_1$, which is
indeed observed in the DNS data as shown in the bottom panel of
Fig.~\ref{fig:E1_vs_epsIN}.  In summary, the asymptotics of the model predicts
qualitative features of the transition which are in agreement with the DNS
results. For further quantitative results, we evaluate the model numerically.

\subsection{Parameter scan for $\nu_1$}
\label{sec:model-scan}
The results of the previous sections demonstrate that the four-scale model is
able to qualitatively reproduce the features of flow states above and below the
critical value of $\nu_1$. In order to obtain the properties of the transition
to a condensate in the model system, we now proceed with a parameter scan. 
The full model given by eqs.~\eqref{eq:evol_econd}-\eqref{eq:evol_ess} is
integrated numerically for each value of $\nu_1$.
\blue{
The values of the coefficients $c_i$, for $1 \leqslant i \leqslant 4$ have been chosen based on 
the values of shell-to-shell transfers \cite{Domaradzki90,Bratanov15}
from DNS data above and below the critical point, 
\begin{align}
\label{eq:c1}
c_1 & = \frac{T(k=k_{\rm LS},p = k_{\rm IN}, q \simeq k_{\rm I})}{\sqrt{\Els}\Ein} = 0.037 \ , \\
\label{eq:c2}
c_2 & = \frac{T(k=k_1,p = k_{\rm IN}, q \simeq k_{\rm IN})}{\sqrt{E_1}\Ein} = 0.043\ , \\
\label{eq:c3}
c_3 & = \frac{T(k=k_1,p = k_{\rm LS}, q \simeq k_{\rm LS})}{\sqrt{E_1}\Els} = 0.0031 \ , \\
\label{eq:c4}
c_4 & = \frac{T(k=k_{\rm SS},p = k_{\rm IN}, q \simeq k_{\rm I})}{\sqrt{E_{\rm SS}}\Ein} = 0.84\ ,
\end{align}
where $k_1 = 1$, $k_{\rm LS}=5$, $k_{\rm IN}=12$ and $k_{\rm SS}=20$.
We choose a cutoff value $E_{1,0}=0.05$, which results in a transition in the interval $0 < \nu_1 < 1$. \\
}

\begin{figure}[H]
\centering
	\includegraphics[width=0.23\textwidth]{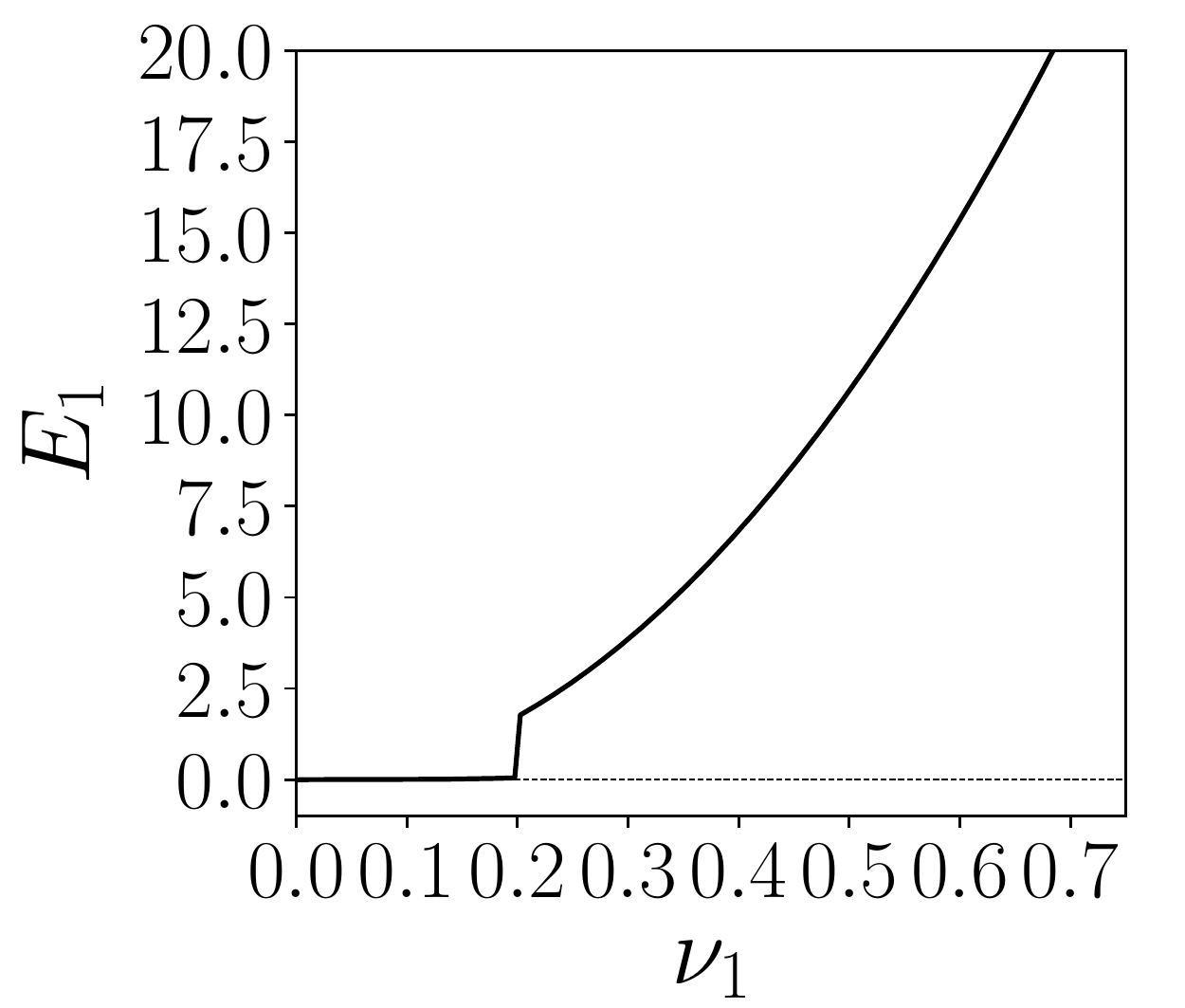} 
	\includegraphics[width=0.23\textwidth]{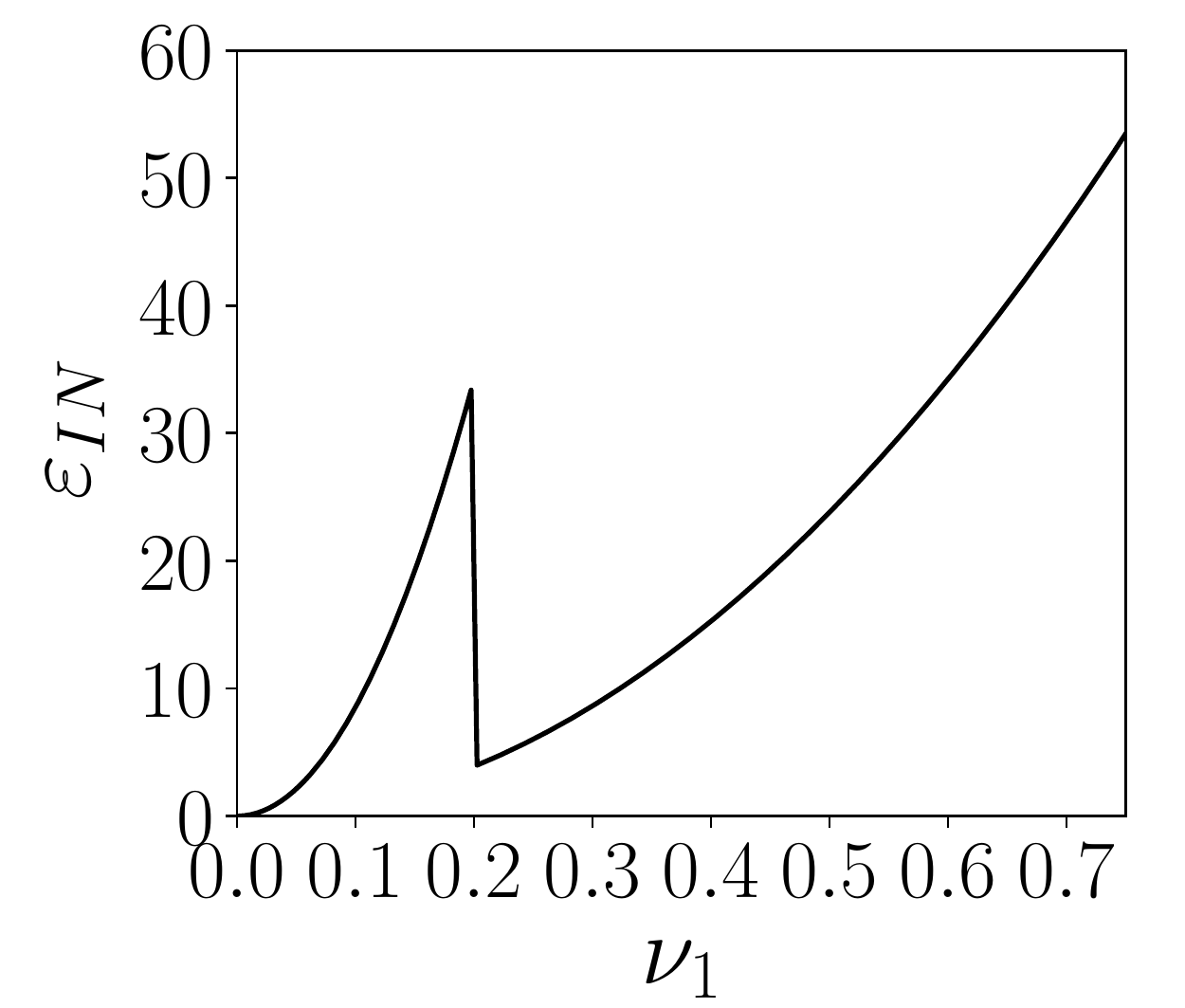} 
	\includegraphics[width=0.23\textwidth]{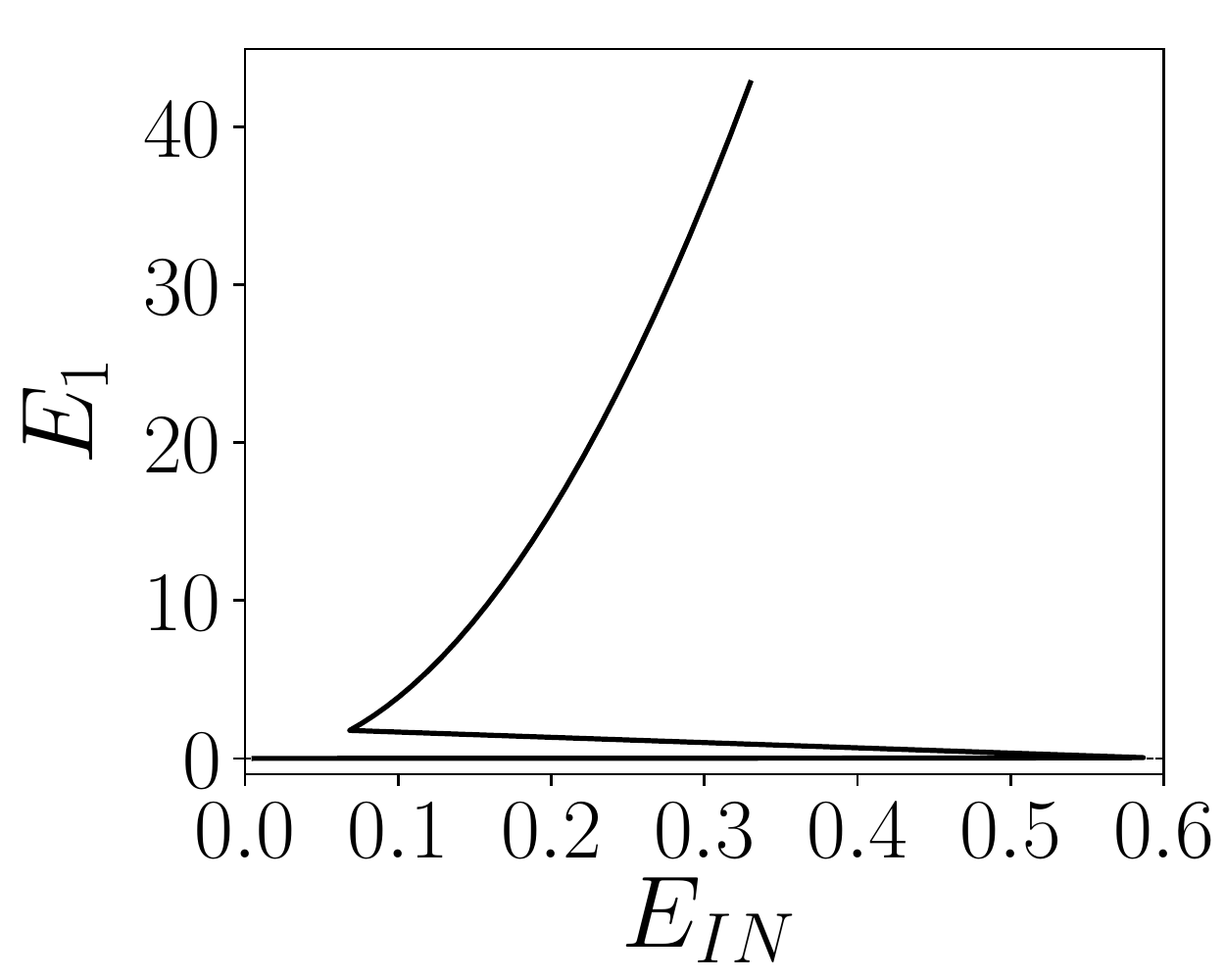} 
	\includegraphics[width=0.23\textwidth]{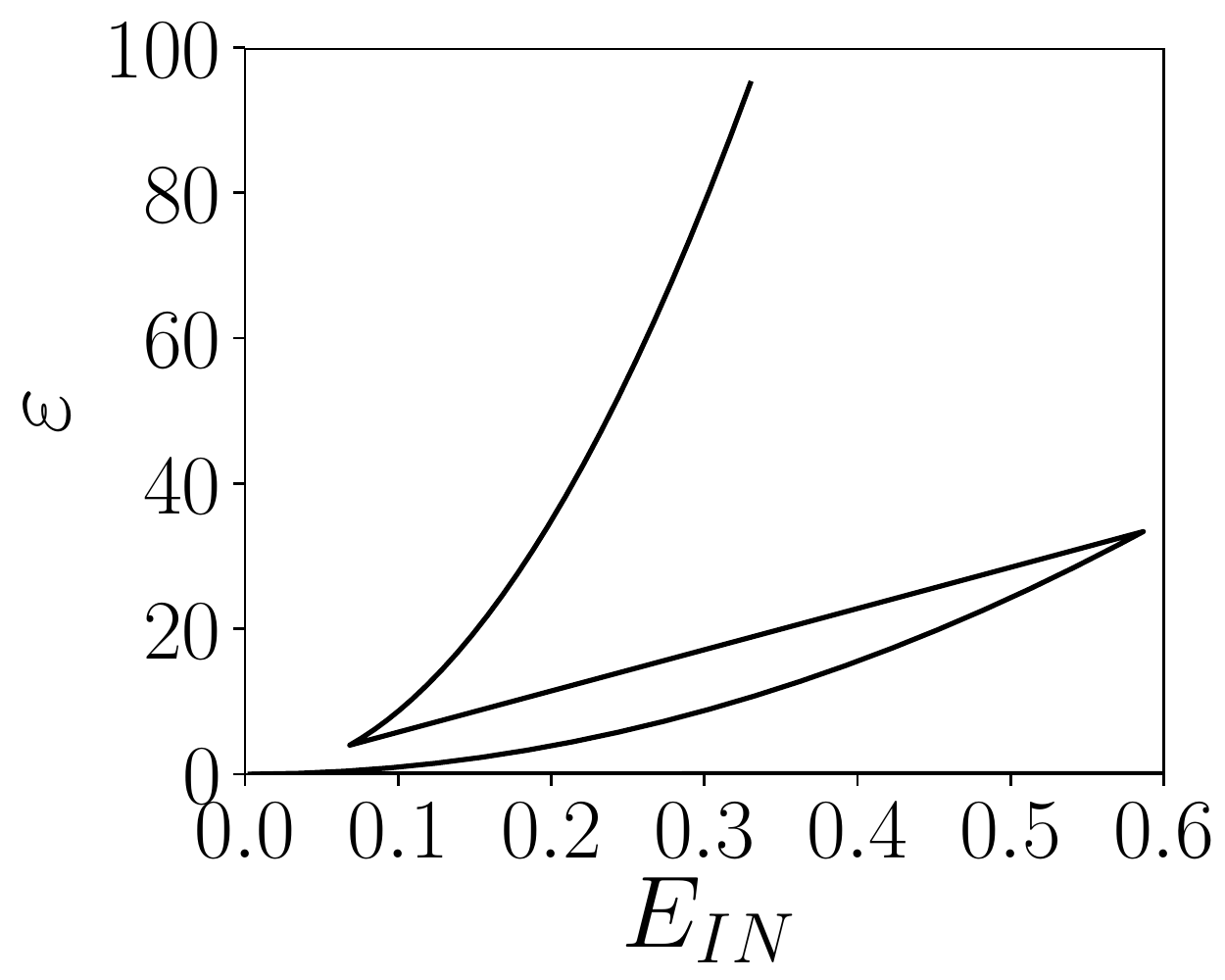} 
 \caption{
	 $E_1$ (left) and $\varepsilon_{\rm IN}$ (right) as functions of $\nu_1$ (top row) and 
	 $\Ein$ (bottom row)
         calculated from a parameter scan of eqs.~\eqref{eq:evol_econd}-\eqref{eq:evol_ess} and
         for $E_{1,0}=0.05$.
         }
 \label{fig:E1_vs_nu1_model}
\end{figure}

A sharp transition must occur in the four-scale model as the dynamics change at
the threshold value $E_{1,0}$ whose qualitative features are remarkably similar
to the transition in the full system.  Figure \ref{fig:E1_vs_nu1_model}
presents the results of the parameter scan for $E_1$ (left panels) and $\epsin$
(right panels) as functions of $\nu_1$ (top row) and $\Ein$ (bottom row). As in the full system, $E_1$ shows a
sudden jump at a critical value of $\nu_1$ and thereafter increases
quadratically in $\nu_1$, while $\Ein$ drops suddenly as predicted for the
asymptotic cases in Secs.~\ref{sec:fourscale-above} and
\ref{sec:fourscale-below}. 

Furthermore, different states of the model system may be realized at the same
value of $\nu_1$, as can be seen from the bottom row of Fig.~\ref{fig:E1_vs_nu1_model}, where
$E_1$ and $\varepsilon$ are presented as functions of $\Ein$.  The sharp
transition is present in form of a discontinuity in the data along a critical
line, and for both $E_1$ and $\varepsilon$ we observe S-shaped curves with
upper and lower branches and an unstable region in between.  This is
qualitatively similar to the behavior of the full system, as can be seen by
comparison with the top panel of Fig.~\ref{fig:E1_vs_epsIN}, which presents the
corresponding DNS data for $E_1(\nu_1/\nu_0)$ and with Fig.~3 of
Ref.~\cite{Linkmann19a} that presents $\eps(\nu_1/\nu_0)$.  We point out that
the model system is not able to track the second, continuous, transition from
absolute equilibrium to viscously damped nonlinear transfers described in
Ref.~\cite{Linkmann19a}, which occurs in the full system at the continuous
inflection point of the lower branch $\varepsilon^-$.  Such an inflection point
is not present in the corresponding model data presented in the lower right panel of
Fig.~\ref{fig:E1_vs_nu1_model}.  This is not surprising as the four-scale
model is by construction not able to produce equipartition of energy between
all degrees of freedom at $k<k_{\rm min}$.

In summary, the model system adequately reproduces the qualitative features of
the transition. The transition is present in the model by construction, where
the model dynamics become nonlocal if a threshold energy at the largest scale
is reached. As such, we suggest that the transition in the full system also
happens through a similar nonlocal coupling scenario: Energy increases at the
largest scales through the classical inverse energy cascade and once a
threshold energy is crossed, the emerging condensate couples directly to the
energy injection range. 

\section{Conclusions}
\label{sec:conclusions}
Active suspensions can be described by a class of one-fluid models that
resemble the Navier-Stokes equation supplemented by active driving provided by
small-scale instabilities originating from active stresses exerted on the fluid
by the microswimmers.  Here, we provided a justification of the one-fluid
approach for the two-dimensional case by relating the solvent's velocity field
non-locally to the coarse grained polarization field of the active
constituents. The resulting model is very similar in structure to solvent
models postulated on phenomenological grounds
\cite{Slomka15EPJST,Slomka17PNAS}. The justification relies on two main
assumptions: The system must be two-dimensional at least to a good
approximation and the bacterial concentration must remain constant. That is,
it is applicable to dense suspensions in thin layers. 

Numerical simulations of a variant of these models showed that a sharp
transition occurs between the formation of a steady-state condensate at the
largest length scale in the system and a steady-state inverse transfer which is
damped by viscous dissipation before reaching the condensate
\cite{Linkmann19a}. The in-depth investigation carried out here supplements the
results of Ref.~\cite{Linkmann19a}, the system is bistable and shows
hysteresis.  That is, 2d active matter turbulence and 2d hydrodynamic
turbulence with a condensate are two non-equilibrium steady states that can
coexist in certain parameter ranges and that are connected through a
subcritical transition.  

The condensate was found to couple directly to the velocity field fluctuations
at the driven scales. This observation led to the introduction of  a
low-dimensional model that includes such a direct nonlinear coupling once a
threshold energy at the largest scales is reached. Analytical and numerical
evaluations of the model resulted in a good qualitative agreement with DNS
results concerning the main features of the transition. As such, we suggest
that the nature of the transition is related to correlations between small- and
large-scale velocity fluctuations. 

Concerning the nature of the transition, we point out that in systems where the
energy input depends on the amount of energy at the driving scales, a reduction
in input occurs at the critical point.  The latter would not be the case for
Gaussian-distributed and $\delta$-in-time correlated random forces as the
time-averaged energy input is known {\em a priori}. In that case, preliminary
results suggest the occurrence of a  supercritical transition (work in progress).
This suggests that the transition to developed
2d-turbulence is highly non-universal: Depending on the type of forcing there
may be no transition, or it may be sub- or supercritical.  Similar situations
occur in rotating flows \cite{Alexakis15,Yokoyama17,Seshasayanan18}.

Several aspects of our results merit further investigation. First and foremost,
it would be of interest to study transitional behavior experimentally. The
Reynolds number necessary for the transition that we found here is at least an
order of magnitude larger than those desccribing mesoscale vortices in dense
bacterial suspensions.  Hence a further increase of swimming speed, a decrease
in viscosity or a larger driving scale are required to trigger the transition.
All three possibilities present considerable difficulty. The most promising
approach may be through the use of non-organic microswimmers.  Second, the
effect of 
friction \mcm{with a substrate}, which is present not only in experiments of
active suspensions but also in the Newtonian case, on the location of the
critical point needs to be quantified.  

\acknowledgments
GB acknowledges financial support by the Departments of Excellence grant
(MIUR).  MCM was supported by the National Science Foundation  through award
DMR-1609208. 

\bibliography{references}

\end{document}